  \documentstyle[aps,eqsecnum,multicol,psfig]{revtex}
  \input epsf
  \begin{document}
\newcommand{\delsh}{{\delta^{\rm s}}}
\newcommand{\shmd}{{E}}
\newcommand{\shex}{{t}}
\newcommand{\cpo}{{W}}
\newcommand{\fft}{\bf}
\newcommand{\wimpf}{\varpi}
\newcommand{\beq}{\begin{equation}}
\newcommand{\eeq}{\end{equation}}
\newcommand{\bea}{\begin{eqnarray}}
\newcommand{\eea}{\end{eqnarray}}
\newcommand{\bml}{\begin{mathletters}}
\newcommand{\eml}{\end{mathletters}}
\newcommand{\ie}{{i.e.}}
\newcommand{\eg}{{e.g.}}
\newcommand{\viz}{{viz}}
\newcommand{\etc}{{etc}}
\newcommand{\real}{{\rm Re\,}}  
\newcommand{\imag}{{\rm Im\,}}  
\newcommand{\eqbreak}{
\end{multicols}
\widetext
\noindent
\rule{.48\linewidth}{.1mm}\rule{.1mm}{.1cm}
}
\newcommand{\eqresume}{
\noindent
\rule{.52\linewidth}{.0mm}\rule[-.1cm]{.1mm}{.1cm}\rule{.48\linewidth}{.1mm}
\begin{multicols}{2}
\narrowtext
}
%
\draft
\title{Semi-microscopic theory of elasticity 
near the vulcanization transition}

\author{Horacio E.~Castillo\rlap,$^{(a)}$ 
and Paul M.~Goldbart$^{(b)}$}
\address{
$^{(a)}$ CNRS-Laboratoire de Physique Th{\'e}orique de l'Ecole
Normale Sup{\'e}rieure,
24 rue Lhomond, 
75231 Paris CEDEX 05, France \\
$^{(b)}$ Department of Physics, University of Illinois at
Urbana-Champaign, 1110 West Green Street, 
Urbana, IL 61801-3080, USA}

  \date{September 3, 1999}

\maketitle
\begin{abstract}

Randomly crosslinked macromolecules undergo a liquid--to--amorphous solid 
phase transition at a critical crosslink concentration.  This transition 
has two main signatures: the random localization of a fraction of the 
monomers and the emergence of a nonzero static shear modulus.  In this 
article, a semi-microscopic statistical mechanical theory of the elastic 
properties of the amorphous solid state is developed.  This theory takes 
into account both quenched disorder and thermal fluctuations, and allows 
for the direct computation of the free energy change of the sample due to 
a given macroscopic shear strain.  This leads to an unambiguous 
determination of the static shear modulus.  At the level of mean field 
theory, it is found (i)~that the shear modulus grows continuously
from zero at the transition, and does so with the classical exponent,
\ie, with the third power of the excess crosslink density and, quite
surprisingly, (ii)~that near the transition the external stresses do
not spoil the spherical symmetry of the localization clouds of the
particles.

\end{abstract}
%
%
\begin{multicols}{2}
\section{Introduction}
\label{SEC:introduction}

The vulcanization transition is the equilibrium phase transition from a
liquid state to a random solid state (known as the amorphous  solid state)
that occurs when a sufficient density of randomly located,  permanent
crosslinking constraints is applied  to the constituents of a liquid.  
The liquid may be a melt of macromolecules of various types (long or
short, flexible or stiff) or even of a low molecular weight species,
and our results will---{\it mutatis mutandis\/}---apply to this broad
variety of systems.  Corrections due to long-wavelength fluctuations
of the order parameter, omitted in the mean field theory that we shall
be developing do, however, tend to be important over a narrower range
of crosslink densities for longer macromolecules and stronger
concentrations of them~\cite{FNOTE:MF_DG}.  For this reason, and for
the sake of 
concreteness, we shall focus on cases involving long, linear, flexible
macromolecules.
There are two main equilibrium signatures of the vulcanization 
transition:  (i)~the structural signature that a nonzero fraction of the
monomers  (i.e.~segments of the macromolecules) become localized around
random  mean positions and have random localization lengths; and (ii)~the
response signature that the system, as a whole, acquires  a nonzero 
static shear modulus. The structural signature has been discussed 
previously; the purpose of this Paper is to present a detailed analysis 
of the the latter signature by developing a statistical-mechanical 
theory of the emergent elastic properties of the amorphous solid state 
in the vicinity of the vulcanization transition.  A core feature of this 
theory, a brief account of which was given in Ref.~\cite{REF:Elas_short}, 
is that it incorporates both annealed (\ie~thermally equilibrating) and 
quenched random (\ie~crosslink specifying) variables.  
Its main conclusions are: 
(a)~that the amorphous state emerging at the vulcanization transition, 
which is solid in the sense of the structural signature~(i), is indeed 
solid in the sense of the response signature~(ii); 
(b)~that the elastic shear modulus
vanishes continuously as the transition is approached, and does so with
the third power of the excess crosslink density (\ie\ the amount by
which the cross-link density exceeds its critical value); 
and (c)~that the shearing of the container associated with elastic 
deformations leads neither to a deterministic nor a stochastic 
shearing of the probability clouds associated with the thermal 
fluctuations of localized particles about their mean positions.

There has been considerable attention paid, over the years, 
to the elastic properties of vulcanized matter and related
chemically-bonded systems, especially those near the amorphous
solidification transition. 
Amongst the most notable approaches are the classical
ones~\cite{REF:DobGor}, in which it was argued that near the
transition the elastic entropy in the solid phase (and consequently
the static shear modulus $\shmd$) grow as the third power of the
excess cross-link density $\epsilon$, \ie,
$\shmd\sim\epsilon^{\shex}$ with $\shex=3$.  
More recently, it was
proposed that the amorphous solidification transition of polymer
systems be identified with a percolation
transition~\cite{REF:PGDGbook,REF:PGDGperc}. 
This proposal led to the identification of the 
exponent $\shex$ with the critical exponent $\mu $ for percolation of
conductivity (with $\mu \approx 2.0$ in $3$ spatial
dimensions). 
In yet more recent work it was observed that the elasticity
percolation exponent for a random network is substantially higher than
$\mu $ when the forces are central~\cite{REF:feng}.
Part of the ambiguity in the determination of the shear modulus of the
randomly crosslinked system from percolative approaches stems from the
fact that these approaches do not naturally lead to the computation of
a free energy for the system.

Approaches of a more microscopic orientation have also been made to 
the elastic properties of vulcanized matter, in which macromolecular 
degrees of freedom feature explicitly. 
Among these are
the ``phantom network''~\cite{REF:phantom} and ``affine
network''~\cite{REF:Florybook} approaches, as well as the comprehensive
discussion of rubber elasticity by Deam and Edwards~\cite{REF:DeamEd},
and others~\cite{REF:Panyukov}.
These approaches focus on the well-crosslinked regime rather than the
lightly-crosslinked regime near the vulcanization
transition~\cite{REF:MarkErman}. 

Experimentally, the exponent $\shex$ has been addressed for several
systems (although mostly for gelation rather than
vulcanization): the results vary from $\shex\approx 2$~\cite{REF:exp2}
to $\shex \agt 3$~\cite{REF:exp3}. 
This wide discrepancy remains unresolved.

The classical~\cite{REF:Flory,REF:Stockmayer,REF:DobGor} and  
percolation~\cite{REF:PGDGbook,REF:PGDGperc} approaches to the
physics of vulcanized matter are certainly stimulating.
However, it must be recognized that neither explicitly includes both 
crucial ingredients: {\em thermal fluctuations\/} and {\em quenched
disorder\/}.  
In addition, as was mentioned above, contradictory results have been
obtained in the determination of the shear modulus of random amorphous
solids from percolative formulations, due in part to the lack of a
natural definition of an elastic free energy for the system.  In the
approach that we shall present, however, the free energy of the system
emerges immediately as a physical quantity, and the value of the shear
modulus is determined unambiguously by the change of the free energy
due to deformations of the sample.
Over the past few years, an approach to the vulcanization transition 
has been developed~\cite{REF:prl_1987,REF:PMGandAZprl,REF:epl,REF:cross} 
that explicitly incorporates both thermal fluctuations and quenched
disorder in the context 
of a semi-microscopic model for flexible, randomly crosslinked 
macromolecules.  
This approach is very much inspired by the work of Edwards and 
collaborators~\cite{REF:DeamEd,REF:Ball}, as well as by ideas 
from the field of spin glasses. 
Emerging from this more recent approach has been a detailed
picture of the {\em structure\/} of the amorphous solid state near to
the vulcanization transition, including, in particular, an explicit
form for the distribution of localization lengths.  
In this Paper, we present a detailed exposition of the application of
this approach to the second signature of the vulcanization transition,
namely the emergence of static {\em response\/} to shear deformations.
To our knowledge, this is the only existing computation of the static
elastic properties of randomly crosslinked macromolecular systems near
the vulcanization transition that starts from first principles 
and thereby includes both the effects of quenched disorder and thermal
fluctuations. 

The outline of the present Paper is as follows. 
This introduction is followed by two long sections. 
In Sec.~\ref{SEC:model} we present the model that we use to describe 
systems of randomly crosslinked macromolecules, and review prior 
results obtained from this model. 
In Sec.~\ref{SEC:response} we describe and implement the changes necessary 
to accommodate strained systems, construct the  appropriate free energy, 
determine the resulting order parameter, and compute the elastic shear 
modulus.  Finally, in Sec.~\ref{SEC:conclusions}, we gives a short summary
of our results.

\section{Model} 
\label{SEC:model}

To help set the stage for our analysis of the elastic response of
a system of randomly crosslinked macromolecules in the amorphous solid
state, in this section we present the model that we use to analyze
systems of randomly crosslinked macromolecules, and review some of the
results about the amorphous solidification transition in those systems
that have been obtained previously within the framework of the same
model. This section is mainly didactic in nature,
specifically aimed to familiarize the reader with the model and the
techniques employed to analyze it, and also to make the present
paper as self contained as possible. In our exposition we mainly
follow Refs.~\cite{REF:prl_1987,REF:PMGandAZprl,REF:epl,REF:cross}, and we
refer the reader interested in further details to these articles
(especially Ref.~\cite{REF:cross}, which gives a detailed account of
the model).

\subsection{Description of the system: macromolecules; Edwards
hamiltonian; random permanent crosslinks} We study a system of $N$
macromolecules of arclength $L$ and persistence length $\ell$ moving
in a $d$-dimensional hypercubic volume $V$. The thermal degrees of
freedom are the positions of the monomers ${\bf c}_{i}(s)$, where the
index $i=1,\dots, N$ labels the macromolecules and the arclength $0\le
s \le 1$ labels the monomers on a given macromolecule. (For
convenience, we measure arclengths in units of the total arclength
$L$, and spatial positions in units of $\sqrt{L\ell/d}$,
\ie\ the r.m.s.~end-to-end distance of a free macromolecule divided by
$\sqrt{d}$.)

We model the system prior to crosslinking by using the Edwards
hamiltonian~\cite{REF:SFE_excv}, 
\begin{eqnarray}
H &=& \frac{1}{2}\sum_{j=1}^{N}\int_{0}^{1} \!\! ds
\left\vert \frac{d{\bf c}_{j}(s)}{{ds}}\right\vert^{2}
\nonumber \\
&& \qquad +\frac{\lambda^{2}}{2}\sum_{i,i^{\prime}=1}^{N}
\int_{0}^{1} \!\! ds\int_{0}^{1} \!\! ds^{\prime}\,
\delta
\big({\bf c}_{i}(s)-{\bf c}_{i^{\prime}}(s^{\prime})\big),
\label{EQ:Edwards_H}
\end{eqnarray}
where $\lambda^{2} (>0)$ characterizes the effect of the (repulsive)
ex{\-}cluded-volume interaction between monomers and $\delta\big({\bf
c}\big)$ is the $d$-dimensional Dirac delta-function.  

We suppose that permanent crosslinks are introduced between a random
number $M$ of randomly selected pairs of monomers: monomer $s_{e}$ on
chain $i_{e}$ is crosslinked to monomer $s_{e}^{\prime}$ on chain
$i_{e}^{\prime}$ (with $e=1,\dots, M$).  These constraints, which
enforce certain pairs of monomers to occupy common spatial locations,
do not break translational symmetry, and the variables that specify
the constraints, $\chi \equiv
\{i_{e},s_{e};i_{e}^{\prime},s_{e}^{\prime}\}_{e=1}^{M}$, play the
role of quenched random variables.
For a particular realization of the
disorder~\cite{FNOTE:entanglements}, the partition function
reads
\begin{equation}
Z\big(\chi \big)
\equiv 
\frac{\bar{Z}\big(\chi \big)}
{\sigma\big(\chi \big)}.
\label{EQ:Z}
\end{equation}
Here, $\bar{Z}\big(\chi \big)$
is a na{\"\i}vely computed sum of the thermodynamic weights for
allowed macromolecular configurations, and is defined via
\begin{equation}
\bar{Z}\big(\chi \big)
\equiv
\!\!\int\!\!{\cal D}{\bf c}\,\,{\rm e}^{-H} \prod_{e=1}^{M}
\!\delta\!\!\left({\bf c}_{i_{e}}(s_{e})-{\bf c}_{i_{e}^{\prime}}
(s_{e}^{\prime})\right), 
\label{EQ:bar_Z}
\end{equation}
where $\int{\cal D}{\bf c}$ indicates functional integration over all
configurations of all macromolecules. The permutation symmetry factor
$\sigma(\chi)$ in Eq.~(\ref{EQ:Z}) depends on the realization of the
disorder and compensates for the overcounting of
configurations that only differ by permutations of the labels of
macromolecules.
If the different types of clusters (\ie, objects composed of
macromolecules attached to each other by crosslinks) present in the
system are labeled by an index $a$, and if $N_{a}$ denotes the number
of identical clusters of type $a$ for the particular disorder
realization $\chi$, then $\sigma(\chi) = \prod_{a} N_{a}!$
\cite{REF:cross}. In particular, in the case of uncrosslinked
macromolecules, which we denote by $\chi = \chi_{0}$, there is only
one cluster type, $N_{a}$ takes the value $N$ for that type, and thus
the permutation factor reduces to the standard value of $1/N!$. The
symmetry factor does not change when the system is deformed after
crosslinking, and is thus irrelevant in the determination of elastic
properties. However, as is the case with the $1/N!$ factor in the
simplest example of all molecules being identical, the symmetry factor
{\em does} play a role in ensuring that the free energy of the system
has the proper extensive scaling.

It should be pointed out that in the present approach the macromolecules 
are allowed to pass through one another and, therefore, the interlocking 
of loops is not explicitly considered~\cite{FNOTE:entanglements}. 
There are reasons to believe that this is a good approximation.
First, the transition regime, of interest here, is characterized 
by a rather low crosslink density---of order one crosslink per 
macromolecule---so most of the macromolecules appear in 
\lq\lq tree-like\rq\rq\ structures and therefore loops might 
reasonably be expected to have little impact.  
Second, under coarse-graining the distinction between holonomic and 
anholonomic constrains tends to fade, with knots and crosslinks having 
rather similar effects. 
Third, a comparison of the results for the gel fraction and the distribution 
of localization lengths obtained from the present approach with those 
obtained in computer simulations that do include the interlocking of 
loops suggests that this interlocking effect is indeed negligible 
near the liquid--to--amorphous solid 
transition~\cite{REF:SJB_MP_a,REF:SJB_MP_b,REF:landau}.
\subsection{Probability distribution of the quenched disorder}
To model the distribution of the crosslink locations in a realistic
vulcanization process, we make the following physical assumption (due
to Deam and Edwards~\cite{REF:DeamEd}): a ``snapshot'' of the
semi-microscopic state of the uncrosslinked system is taken, and, if a
pair of monomers happen to be close to each other, there is a 
probability, determined by a parameter $\mu ^2$, of becoming permanently
attached by a crosslink. This translates into the crosslink
distribution~\cite{REF:DeamEd}
\begin{eqnarray}
&&
{\cal P}_{M} 
\big(\chi \big)
\frac{{\cal C}}{M!} \left(\frac{\mu ^{2}V}{2N}\right)^{M} 
\tilde{Z}\big(\chi \big),
\label{EQ:DE_dist_def}
\end{eqnarray}
where $\tilde{Z}$ is a correlator that probes the statistics of
contact points between macromolecules,\ie,
\begin{equation}
\tilde{Z}\big(
\chi \big)
\equiv 
\left\langle
\prod_{e=1}^{M}
\delta\big({\bf c}_{i_{e}}(s_{e})-{\bf c}_{i_{e}^{\prime}}
(s_{e}^{\prime})\big)
\right\rangle_{1}^{\rm E},  
\label{EQ:DE_dist_corr}
\end{equation}
and ${\cal C}$ is a normalization factor, given by
\begin{equation}
{\cal C}^{-1} = \Big\langle \exp\Big\{\frac{\mu ^{2}V}{2N}
\sum_{i,j=1}^{N} \int_{0}^{1} \!\! ds\int_{0}^{1} \!\! dt\,
\delta
\left({\bf c}_{i}(s)-{\bf c}_{j}(t)\right) \Big\} 
\Big\rangle_{1}^{\rm E}.
\label{EQ:DE_dist_norm}
\end{equation}
The angle brackets denote normalized averaging with respect to one
copy of the Edwards ham\-il\-to\-ni\-an:
\begin{equation}
\langle\cdots\rangle_{1}^{\rm E}\equiv~
\frac{\int{\cal D}{\bf c}\,{\rm e}^{-H} \cdots} {\int{\cal D}{\bf c}\,{\rm
e}^{-H}}.
\label{EQ:EAv_def}
\end{equation}
The correlator $\tilde{Z}(\chi)$ is proportional to the probability of
finding a particular set of pairs of monomers in close proximity. It
can also be interpreted as the ratio $\tilde{Z}(\chi) = \bar{Z}(\chi)
/ \bar{Z}(\chi_{0})$ between two na{\"\i}ve partition functions: the
numerator for the given set of crosslinks $\chi$ and the denominator
for the case of uncrosslinked macromolecules $\chi_{0}$.

With the distribution ${\cal P}$, the mean number of crosslinks per
macromolecule $[M]/N$ is a smooth, monotonically-increasing function
of the control parameter $\mu ^{2}$. Here and subsequently, square
brackets $[\cdots]$ denote averages over the crosslink distribution,
\ie\ disorder averaging.

\subsection{Order parameter for amorphous solidification}

Let us now, in the spirit of the Landau theory of phase transitions, 
review the definition of an order parameter that captures the
distinctions between the possible states of a randomly crosslinked
macromolecular systems~\cite{REF:prl_1987}.

Consider the real-space probability density $\rho_{j,s,\chi}({\bf r})$
for one particular monomer $(i,s)$ to be at position ${\bf r}$ in the sample,
\begin{equation}
\rho_{j,s,\chi}({\bf r}) \equiv \langle \delta({\bf r} - {\bf c}_{j}(s))
\rangle_{\chi}, 
\label{EQ:density_r_def}
\end{equation} 
as well as its Fourier transform
\begin{equation}
\langle \exp(i {\bf k \cdot c}_{j}(s)) \rangle_{\chi} \equiv \int \!
d{\bf r} \, \exp(i {\bf k \cdot r}) \rho_{j,s,\chi}({\bf r}), 
\label{EQ:density_k_def}
\end{equation} 
where ${\bf k}$ is any wave vector.
The angle brackets $\langle\cdots\rangle_{\chi}$ indicate an average
over the equilibrium state in question for a particular realization
$\chi$ of the disorder, as we indicate by the subscript $\chi$.  If
the monomer is delocalized, its density is uniform in real space, and
is a Kronecker $\delta$-function centered at ${\bf k = 0}$ in Fourier
space. On the other hand, if the monomer is localized in the vicinity
of a certain point, say ${\bf b}_{j}(s)$, then the density in Fourier
space is no longer a Kronecker delta function, and one instead expects
it to have the form~\cite{REF:cross}
\begin{equation}
\langle \exp(i {\bf k \cdot c}_{j}(s)) \rangle_{\chi} = \exp(i {\bf k
\cdot b}_{j}(s)) \, {\cal F}_{j,s,\chi}({\bf k}),
\label{EQ:individual_density}
\end{equation} 
where the factor ${\cal F}_{j,s,\chi}({\bf k})$ is a form factor that
describes the thermal fluctuations of the monomer about its average
position.

Now, one could be simply tempted to propose the disorder average of
the total density (\ie\ the average of all individual monomer
densities) as the order parameter. In fact, this kind of order
parameter would allow one to distinguish between a liquid phase, in
which all elements of the system are delocalized, and a crystalline
phase, in which the system forms some kind of regular lattice in real
space, or a globule phase, in which the density is concentrated in one
particular region in real space (see Table \ref{TAB:states}).
However, an amorphous solid phase, in which some of the monomers are
localized, but in a random and homogeneous manner, would give the same
value of the order parameter as a liquid (except for corrections of
subleading order in the thermodynamic limit). Thus, we need an order
parameter that probes the structure of the system in a subtler way.
A similar problem appears in the theory of spin glasses, in which a
system of $N$ spins $\{{\bf S}_{j}\}_{j=1}^{N}$ are subject to random
frustrated interactions. The total magnetization ${\bf M} = (1/N)
\sum_{j=1}^{N} \langle {\bf S}_{j} \rangle $  (\ie\ the sum of the thermal
averages of all the spins in the system) is zero both in the
paramagnetic phase, where each term in the sum is zero, and in the
spin glass phase, in which individual terms in the sum are nonzero but
cancel each other because of their random orientations.  The
solution to this difficulty in the spin glass case was found by
Edwards and Anderson~\cite{REF:SFEandPWA}: one should take a
sum of {\em products} of the local mean values: $(1/N) \sum_{j=1}^{N}
\langle {\bf S}_{j} \rangle \langle {\bf S}_{j} \rangle$, and the
cancellation of terms no longer occurs. 
A related approach also works in
the context of the vulcanization transition, where the order parameter is
\cite{REF:prl_1987}:
\begin{equation}
\Omega_{{\bf k}^{1},
\ldots,{\bf k}^{g}}
\equiv
\left[
\frac{1}{N}\sum_{j=1}^{N}\int_{0}^{1}ds\,
\langle {\rm e}^{i{\bf k}^{1}\cdot {\bf c}_{j}(s)} \rangle_{\chi}
\cdots
\langle {\rm e}^{i{\bf k}^{g}\cdot {\bf c}_{j}(s)} \rangle_{\chi}
\right].
\label{EQ:opDefinition}
\end{equation}
Here, $g$ is a positive integer, and $\{{\bf k}^{1},\ldots,{\bf
k}^{g}\}$ are any $g$ non-zero $d$-dimensional wave vectors. The order
parameter $\Omega_{{\bf k}^{1},\cdots,{\bf k}^{g}}$ is, in principle,
experimentally accessible through scattering experiments: \eg\ the
incoherent contribution to the elastic neutron scattering
cross-section is proportional to $\Omega_{{\bf k},{\bf -k}}$ 
\cite{REF:cross,REF:Mezei}.

In the liquid state, all terms in the sum in
Eq.~(\ref{EQ:opDefinition}) are zero, and the order parameter is zero.
In the amorphous solid state, a nonzero fraction of the monomers are
localized in the vicinity of certain points in space, and one expects
their individual densities to have the form given by
Eq.~(\ref{EQ:individual_density}).  In this case, there are terms in
the sum of Eq.~(\ref{EQ:opDefinition}) that are nonzero, and we only
need to avoid their possible mutual cancellation due to their random
phases. This is readily done in the case $g>1$ by choosing the wave
vectors so that ${\bf k}^{1}+\cdots+{\bf k}^{g}={\bf 0}$.

\eqbreak
\begin{table}[htb]
\caption{States of the system, order parameter values, and
symmetries.}
\begin{tabular}{cccc}
State	& Density for 
	& Order parameter 
	& Translational 
\\
	& one monomer 
	& $\Omega_{{\bf k}^{1},\ldots,{\bf k}^{g}}$ 
	& symmetries 
	\\ 
\tableline
liquid	& $ \langle {\rm e}^{i {\bf{k \cdot r}}_{j}}\rangle = \delta_{\bf{k, 0}}$ (all) 
	& $\delta_{{\bf k}^{1}, \bf{0}} \times \cdots \times \delta_{{\bf k}^{g}, \bf{0}} $	
	& macroscopic and 
\\ 
	&	
	&	
	& microscopic 
\\ 
\tableline
amorphous & $ \langle {\rm e}^{i {\bf{k \cdot r}}_{j}}\rangle 
		\propto {\rm e}^{i {\bf{k \cdot b}}_{j}} $ (some) 
	& $\propto \delta_{{\bf k}^{1}+\cdots+{\bf k}^{g}, {\bf 0}} $ 
	& macroscopic 
\\ 
solid	& $ \langle {\rm e}^{i {\bf{k \cdot r}}_{j}}\rangle = \delta_{\bf{k, 0}}$ (others) 
	&	& 	
\\ 
\tableline
crystal	& $ \langle {\rm e}^{i {\bf{k \cdot r}}_{j}}\rangle 
		\propto {\rm e}^{i {\bf{k \cdot b}}_{j}}$ 
	& $\propto \delta_{{\bf k}^{1}+\cdots+{\bf k}^{g}, {\bf G}} $ 
	& macroscopic 
\\
	& 
	& $ \bf{G} \in $ reciprocal lattice	
	& (only by discrete	
\\
	&	
	&	
	& lattice vectors) 
\\ 
\tableline
globule	
	& $ \langle {\rm e}^{i {\bf{k \cdot r}}_{j}}\rangle 
		\propto {\rm e}^{i {\bf{k \cdot b}}_{j}} $ (some) 
	& $ S({\bf k}^{1}+\cdots+{\bf k}^{g})$	
	& none	
\\ 
(with density $\rho(\bf{r})$) 
	& $ \langle {\rm e}^{i {\bf{k \cdot r}}_{j}}\rangle = \delta_{\bf{k, 0}} $ (others) 
	& ($S(\bf{k}) \equiv \int d{\bf k}\, {\rm e}^{i {\bf{k \cdot r}}} \rho(\bf{r})$)  
	& 	
\\
\end{tabular}
\label{TAB:states}
\end{table}
\eqresume
{}From the point of view of symmetry, states that have a nonuniform
average total particle density, such as the crystalline and globular
states, break translation invariance macroscopically. On the other
hand, states with a uniform average total particle density, such as the
liquid and the amorphous solid, are macroscopically translation
invariant (MTI). If we now look at the individual monomer densities,
they are only spatially uniform for all monomers for the case of the
liquid state, and thus we say that the liquid state is microscopically
translationally invariant, but that is not the case for the amorphous
solid, the crystalline solid or the globule.  It can be seen, \eg, by
applying a common displacement to all monomers, that the order
parameter for an MTI state can only be nonzero when 
${\bf k}^{1}+\cdots+{\bf k}^{g}={\bf 0}$. 
Table~\ref{TAB:states} summarizes the above discussion.

For the sake of computational simplicity, we adopt periodic boundary
conditions on the coordinates to describe the microscopic
configurations of our system. Consequently, the order parameter is
only nonzero for ${\bf k}^{a} \in r^{\rm u}, a = 1,\ldots,g $, where
$r^{\rm u}$ denotes the reciprocal lattice in $d$-dimensions
associated with the periodicity in real space arising from the
boundary conditions.

\subsection{Averages over the quenched disorder: the replica technique}
\label{SEC:av_rep}
To make progress analytically, we need to be able to compute averages
over the quenched disorder associated with the random crosslink
locations, \eg, the disorder-averaged free energy $f$ per
macromolecule
\begin{equation}
f \equiv \frac{1}{N} [\ln{Z}] = \frac{1}{N} [\ln{\bar{Z}}] -
\frac{1}{N} [\ln{\sigma}]. 
\label{EQ:f_def}
\end{equation}
The second term on the extreme right hand side of Eqn.~(\ref{EQ:f_def})
plays no role in the elastic properties of the system, as mentioned
above. (It does not even play any role in determining the order
parameter.)
For this reason we henceforth ignore it, and
focus instead on the ``na{\"\i}ve free energy'' (per macromolecule)
\begin{equation}
\bar{f} \equiv \frac{1}{N} [\ln{\bar{Z}}],
\label{EQ:bar_f_def}
\end{equation}
which is adequate for our purposes.
In order to compute the average over disorder, we
make use of the replica technique. Within the framework of this
technique, $\bar{f}$ is obtained from the prescription
\begin{equation}
\bar{f} = \lim_{n \to 0} \frac{[{\bar{Z}}^{n}] -1}{n N},
\label{EQ:f_Z}
\end{equation}
where 
${\bar{Z}}^{n}$ is interpreted, as usual, as the partition
function for a system comprising $n$ identical copies (or {\em
replicas}) of the original system. 

For the Deam-Edwards distribution, Eq.~(\ref{EQ:DE_dist_def}), the
disorder average in Eq.~(\ref{EQ:f_Z}) takes the form
\begin{equation}
[{\bar{Z}}^{n}] = {\cal C} \sum_{\chi} \frac{1}{M_{\chi}!}
\left(\frac{\mu ^{2}V}{2N}\right)^{M_{\chi}} \tilde{Z}\big(\chi \big)
\, ({\bar{Z}}\big(\chi \big))^{n},  
\label{EQ:Z_n_n1}
\end{equation}
where $M_{\chi}$ is the number of crosslinks for the disorder
realization $\chi$. 
The presence of the $\tilde{Z}$ factor coming from the crosslink
distribution 
introduces an
additional replica, besides the $n$ replicas associated with the $n$
factors of ${\bar{Z}}$.
The additional replica, which we label by $\alpha=0$, represents
the degrees of freedom of the original system before crosslinking,
or, equivalently, encodes the consequences of the cross-link distribution. Consequently,
any external strain applied to the system {\em after the permanent
constraints have been created} will affect replicas
$\alpha=1,\ldots,n$, but not replica $\alpha=0$~\cite{REF:DeamEd}.
\eqbreak

\subsubsection{Computing $[\bar{Z}^{n}]$ explicitly: particle density
variables} 
A more explicit expression for $[\bar{Z}^{n}]$ follows from the
combining of 
Eqs.~(\ref{EQ:bar_Z}), (\ref{EQ:DE_dist_corr}), (\ref{EQ:DE_dist_norm})
and (\ref{EQ:Z_n_n1}):
\begin{eqnarray}
[\bar{Z}^{n}] & = & \frac{
\int {\cal D}{\bf c}^{0}\cdots{\cal D}{\bf c}^{n} 
\,\,\,{\rm e}^{-\sum_{\alpha=0}^{n} H^{\alpha}} 
\sum_{\chi} \frac{1}{M_{\chi}!}
\left(\frac{\mu ^{2}V}{2N}\right)^{M_{\chi}} 
\prod_{\alpha=0}^{n} \prod_{e=1}^{M_{\chi}} 
\delta\left({\bf c}^{\alpha}_{i_{e}}(s_{e})-{\bf
c}^{\alpha}_{i_{e}^{\prime}} (s_{e}^{\prime})\right) }
{ 
\int {\cal D}{\bf c} \,\,\, \exp\big\{ -H + \frac{\mu ^{2}V}{2N} 
\sum_{i,j=1}^{N} \int_{0}^{1} \!\! ds\int_{0}^{1} \!\! dt\, 
\delta \left({\bf c}_{i}(s)-{\bf c}_{j}(t)\right) \big\} 
}
\nonumber \\
\noalign{\medskip}
& = & \frac{
\int {\cal D}\hat{c}
\,\,\,\exp\Big\{-\sum_{\alpha=0}^{n} H^{\alpha} + \frac{\mu ^{2}V}{2N} 
\sum_{i,j=1}^{N} \int_{0}^{1} \!\! ds\int_{0}^{1} \!\! dt\, 
\delta \left({\hat c}_{i}(s)-{\hat c}_{j}(t)\right) 
\Big\} }
{ 
\int {\cal D}{\bf c} \,\,\, \exp\big\{ -H + \frac{\mu ^{2}V}{2N} 
\sum_{i,j=1}^{N} \int_{0}^{1} \!\! ds\int_{0}^{1} \!\! dt\, 
\delta \left({\bf c}_{i}(s)-{\bf c}_{j}(t)\right) \big\} 
}.
\label{EQ:Z_n_expl}
\end{eqnarray} 
Here, quantities with superscripts $\alpha$ ($=0, 1, \ldots, n$) are
associated with replicas $\alpha$; hatted vectors denote
replicated collections of vectors, viz., ${\hat{v}}\equiv\{{\bf
v}^{0},{\bf v}^{1},\ldots,{\bf v}^{n}\}$, their scalar product being
${\hat{v}}\cdot {\hat{w}}\equiv \sum_{\alpha=0}^{n}{\bf
v}^{\alpha}\cdot {\bf w}^{\alpha}$; the functional integral $\int
{\cal D}\hat{c}$ is equivalent to $\int {\cal D}{\bf c}^{0}\cdots{\cal
D}{\bf c}^{n}$; and we define the Dirac delta function for replicated
vectors via $\delta({\hat{v}}) \equiv \prod_{\alpha=0}^{n} \delta({\bf
v}^{\alpha})$.  Later we shall use the terms {\it one-replica
sector\/} and {\it higher-replica sector\/} to refer to replicated
wave vectors $\hat{k}$ having, respectively, exactly one and more than
one replica $\alpha$ for which the corresponding vector ${\bf
k}^{\alpha}$ is nonzero.
\eqresume
At this point it is convenient to switch from coordinates representing
the macromolecular configurations ${\bf c}_{j}(s)$ and ${\hat c}_{j}(s)$ to
variables representing 
Fourier-transformed monomer densities, defined by
\bml
\begin{eqnarray}
Q_{\bf p}  & \equiv & \frac{1}{N} \sum_{j=1}^{N} \int_{0}^{1} \!\!ds\,\,
{\rm e}^{i {\bf p} \cdot {\bf c}_{j}(s) },
\label{EQ:Q1_def}
\\
Q_{\hat p} & \equiv & \frac{1}{N} \sum_{j=1}^{N} \int_{0}^{1} \!\!ds\,\,
{\rm e}^{i {\hat p} \cdot {\hat c}_{j}(s) }.
\label{EQ:Qh_def}
\end{eqnarray}
\eml
We start by using the decompositions of the Dirac delta function 
in terms of plane waves appropriate for periodic boundary conditions,
and given by 
\bml
\begin{equation}
\delta({\bf c}) = \frac{1}{V} \sum_{\bf p} \exp\left(i{\bf p}\cdot {\bf
c}\right)
\label{EQ:delta_unrep}
\end{equation}
in $d$-dimensional space, and by 
\begin{equation}
\delta({\hat c}) = \frac{1}{V^{1+n}} \sum_{\hat p} \exp\left(i{\hat
p}\cdot {\hat c}\right)
\label{EQ:delta_rep}
\end{equation}
\eml
in replicated space, which allow us to write the crosslink generated
terms in Eq.~(\ref{EQ:Z_n_expl}) as 
\begin{eqnarray} 
\frac{\mu ^{2} V}{2 N} \sum_{i,j=1}^{N} \int_{0}^{1} \!\! ds\int_{0}^{1}
\!\! dt\, \delta \Big({\bf c}_{i}(s)-{\bf c}_{j}(t)\Big) 
& = &
\frac{\mu ^{2} N}{2} \sum_{\bf p} \vert Q_{\bf p} \vert^{2}
\nonumber \\
\label{EQ:int_Q1}
\\
\frac{\mu ^{2} V}{2 N} \sum_{i,j=1}^{N} \int_{0}^{1} \!\! ds\int_{0}^{1}
\!\! dt\, \delta \Big({\hat c}_{i}(s)-{\hat c}_{j}(t)\Big) 
& = &
\frac{\mu ^{2} N}{2 V^{n}} \sum_{\hat p} \vert Q_{\hat p} \vert^{2}.
\nonumber \\
\label{EQ:int_Qh}
\end{eqnarray}

Throughout this calculation we employ periodic boundary conditions in
real space, and this determines the set $r^{\rm u}$ of allowed
$d$-dimensional wave vectors ${\bf p}$ that enter the sums on the
right hand sides of Eqs.~(\ref{EQ:delta_unrep}) and
(\ref{EQ:int_Q1}). The set $R^{\rm u}$ of allowed replicated wave
vectors $\hat{p}$ on the right hand sides of Eqs.~(\ref{EQ:delta_rep})
and (\ref{EQ:int_Qh}) is obtained by taking all combinations of
$(n+1)$ allowed $d$-dimensional wave vectors. Here, the superscript
$\rm u$ stands for ``unstrained system''. In the next section,
when we discuss deformations of the system, these deformations will
directly change the boundary conditions in real space and,
consequently, the set of allowed wave vectors.
\eqbreak

We now can rewrite the expression of 
Eq.~(\ref{EQ:Z_n_expl}) for $[\bar{Z}^{n}]$ in terms of
Fourier transformed monomer densities, 
\begin{equation}
[\bar{Z}^{n}] 
= 
\frac{
{\rm e}^{-Nn \phi}
\int {\cal D}{\hat c}
\,\,\,\exp\Big\{-\frac{1}{2}\sum_{j=1}^{N}\int_{0}^{1} \!\! ds
\left\vert \frac{d{\hat c}_{j}(s)}{{ds}}\right\vert^{2}
- N \tilde{\lambda}_{n}^{2} \frac{N}{V}
\tilde{\sum}^{\dagger}_{\hat{p}} \vert Q_{\hat p} \vert^{2} 
+ N \frac{\mu ^{2}}{V^{n}} {\overline{\sum}}^{\dagger}_{\hat{k}}  
\vert Q_{\hat k} \vert^{2} 
\Big\} }
{ 
\int {\cal D}{\bf c} \,\,\,
\exp\big\{-\frac{1}{2}\sum_{j=1}^{N}\int_{0}^{1} \!\! ds 
\left\vert \frac{d{\bf c}_{j}(s)}{{ds}}\right\vert^{2} -
N \tilde{\lambda}_{0}^{2} \frac{N}{V}
\sum^{\dagger}_{\bf p} \vert Q_{\bf p} \vert^{2}
\big\} 
}.
\label{EQ:Z_n_Q}
\end{equation}
Here, $\phi \equiv \frac{N}{2 V} \lambda^{2} + \frac{\mu ^{2}}{2}
\ln{V} + {\cal O}(n)$ and $\tilde{\lambda}_{n}^{2} \equiv \lambda^{2} -
\mu ^{2} \frac{V}{N} \frac{1}{V^{n}} $ are constants, the symbol
$\tilde{\sum}_{\hat{p}}$ denotes a sum over replicated wave vectors in
the one replica sector, and the symbol $\overline{\sum}_{\hat{k}}$
denotes a sum over replicated wave vectors in the higher replica
sector.  The $\dagger$ symbol additionally restricts any summation to
the half-space of $d$-dimensional [$(n+1)d$-dimensional replicated]
wave vectors such that their scalar product with a fixed unit vector
${\bf n}$ (${\hat n}$) is positive.

\subsubsection{Computing the order parameter}
Having obtained a more explicit expression for
$[\bar{Z}^{n}]$, it is instructive to do the same with the order
parameter. The order parameter is a sum of terms of the form $[
\langle {\cal O}_{1} \rangle_{\chi}  \cdots  \langle {\cal
O}_{g} \rangle_{\chi} ]$, \ie, disorder averages of products of
thermal averages of observables ${\cal O}_{i}$ ($i=1,\ldots,g$). As 
$\langle {1} \rangle_{\chi} = 1$ for any disorder realization $\chi$,
it is evident that
\begin{eqnarray}
\lefteqn{ \left[ \langle {\cal O}_{1} \rangle_{\chi}  \cdots  \langle {\cal
O}_{g} \rangle_{\chi} \right] 
= \lim_{n \to 0} 
\left[ \langle {\cal O}_{1} \rangle_{\chi}  \cdots  \langle {\cal
O}_{g} \rangle_{\chi} 
\langle {1}_{g+1} \rangle_{\chi}  \cdots  \langle {1}_{n}
\rangle_{\chi} \right] }   
\nonumber \\
& = & \lim_{n \to 0}
\left[ 
\frac{ 
\int {\cal D}{\bf c}_{1} {\rm e}^{-H^{1}} \Delta_{1}(\chi) {\cal O}_{1} 
}{
\int {\cal D}{\bf c}_{1} {\rm e}^{-H^{1}} \Delta_{1}(\chi)   
}  
\cdots 
\frac{ 
\int {\cal D}{\bf c}_{g} {\rm e}^{-H^{g}} \Delta_{g}(\chi) {\cal O}_{g} 
}{
\int {\cal D}{\bf c}_{g} {\rm e}^{-H^{g}} \Delta_{g}(\chi)   
}  
\frac{ 
\int {\cal D}{\bf c}_{g+1} {\rm e}^{-H^{g+1}} \Delta_{g+1}(\chi) 
}{
\int {\cal D}{\bf c}_{g+1} {\rm e}^{-H^{g+1}} \Delta_{g+1}(\chi)   
}  
\cdots 
\frac{ 
\int {\cal D}{\bf c}_{n} {\rm e}^{-H^{n}} \Delta_{n}(\chi) 
}{
\int {\cal D}{\bf c}_{n} {\rm e}^{-H^{n}} \Delta_{n}(\chi)   
}  
\right]
\nonumber \\
& = & \lim_{n \to 0}
\left[ 
\int \!\!{\cal D}{\bf c}_{1} {\rm e}^{-H^{1}} \Delta_{1}(\chi) \, {\cal
O}_{1}  
\cdots 
\int \!\!{\cal D}{\bf c}_{g} {\rm e}^{-H^{g}} \Delta_{g}(\chi) \, {\cal
O}_{g}
\int \!\!{\cal D}{\bf c}_{g+1} {\rm e}^{-H^{g+1}} \Delta_{g+1}(\chi) 
\cdots 
\int \!\!{\cal D}{\bf c}_{n} {\rm e}^{-H^{n}} \Delta_{n}(\chi) 
\right]. 
\label{EQ:op_av}
\end{eqnarray}
Here, we have used the definition of the thermal average $\langle {\cal
O} \rangle_{\chi} \equiv \int {\cal D}{\bf c} \,{\rm e}^{-H}
\!\Delta(\chi) {\cal O} / \int {\cal D}{\bf c} \,{\rm
e}^{-H} \!\Delta(\chi) $ and we have denoted by $\Delta(\chi)$ a
quantity that implements the constraints, \ie, that is nonzero only
for those configurations that satisfy the constraints denoted by $\chi$.

Note that at this point the problematic $\chi$-dependent denominators
have been eliminated. We are now able to perform the disorder average
explicitly, thus obtaining an expression analogous to
Eq.~(\ref{EQ:Z_n_expl}). (When computing $\Omega_{{\bf
k}^{1},\cdots,{\bf k}^{g}}$ in the replica approach, both here and
later, we choose $\hat{l}$ so that ${\bf l}^{\alpha} = {\bf
k}^{\alpha}$ for $\alpha = 1, \ldots, g$ and ${\bf l}^{\alpha} = {\bf
0}$ for $\alpha = 0$ and $\alpha = g+1, \cdots, n$),
\begin{eqnarray}
\lefteqn{ \Omega_{{\bf k}^{1},\ldots,{\bf k}^{g}} 
= \lim_{n \to 0} \frac{1}{N} \sum_{j=1}^{N} \int_{0}^{1} \!\! ds
\left[
\langle {\rm e}^{i{\bf k}^{1}\cdot {\bf c}_{j}(s)} \rangle_{\chi}
\cdots
\langle {\rm e}^{i{\bf k}^{g}\cdot {\bf c}_{j}(s)} \rangle_{\chi}
\langle {1}_{g+1} \rangle_{\chi}  \cdots  \langle {1}_{n}
\rangle_{\chi} \right]
} \nonumber \\
& = & \lim_{n \to 0} \frac{1}{N} \sum_{j=1}^{N} \int_{0}^{1} \!\! ds
\left[ \prod_{\alpha=0}^{n} \langle {\rm e}^{i{\bf l}^{\alpha}\cdot {\bf
c}_{j}(s)} \rangle_{\chi} \right] 
\nonumber \\
& = & \lim_{n \to 0}  
\frac{
\int {\cal D}{\bf c}^{0}\cdots{\cal D}{\bf c}^{n} 
\, \left\{ \frac{1}{N} \sum_{j=1}^{N} \int_{0}^{1} \!\! ds \,\,{\rm
e}^{i\sum_{\alpha=0}^{n} {\bf l}^{\alpha} \cdot {\bf
c}^{\alpha}_{j}(s)} \right\}
\,{\rm e}^{-\!\sum_{\alpha=0}^{n} \!H^{\alpha}} 
\sum_{\chi} \frac{1}{M_{\chi}!}
\!\left(\frac{\mu ^{2}V}{2N}\right)^{\!M_{\chi}} 
\prod_{\alpha=0}^{n} \prod_{e=1}^{M_{\chi}} 
\delta\!\left({\bf c}^{\alpha}_{i_{e}}(s_{e})-{\bf
c}^{\alpha}_{i_{e}^{\prime}} (s_{e}^{\prime})\right) }
{ 
\int {\cal D}{\bf c}
\,\,\, \exp\big\{ -H
+ \frac{\mu ^{2}V}{2N}  
\sum_{i,j=1}^{N} \int_{0}^{1} \!\! ds\int_{0}^{1} \!\! dt\, 
\delta \left({\bf c}_{i}(s)-{\bf c}_{j}(t)\right) \big\} 
}
\nonumber \\
& = & \lim_{n \to 0}
\frac{
\int {\cal D}\hat{c} \,\, Q_{\hat{l}}
\,\, \exp\Big\{-\sum_{\alpha=0}^{n} H^{\alpha} + \frac{\mu ^{2}V}{2N} 
\sum_{i,j=1}^{N} \int_{0}^{1} \!\! ds\int_{0}^{1} \!\! dt\, 
\delta \left({\hat c}_{i}(s)-{\hat c}_{j}(t)\right) 
\Big\} }
{ 
\int {\cal D}{\bf c}
\,\,\, \exp\big\{ -H
+ \frac{\mu ^{2}V}{2N}  
\sum_{i,j=1}^{N} \int_{0}^{1} \!\! ds\int_{0}^{1} \!\! dt\, 
\delta \left({\bf c}_{i}(s)-{\bf c}_{j}(t)\right) \big\} 
}.
\label{EQ:Omega_expl}
\end{eqnarray} 

Clearly, the value of the last expression is unchanged by dividing it by
$1 = [ \langle {1}_{1} \rangle_{\chi}  \cdots  \langle
{1}_{g} \rangle_{\chi} ]$. This leads to 
\beq
\Omega_{{\bf k}^{1},\cdots,{\bf k}^{g}} 
=
\lim_{n \to 0}
\frac{
\int {\cal D}\hat{c} \,\, Q_{\hat{l}}
\,\, \exp\Big\{-\sum_{\alpha=0}^{n} H^{\alpha} + \frac{\mu ^{2}V}{2N} 
\sum_{i,j=1}^{N} \int_{0}^{1} \!\! ds\int_{0}^{1} \!\! dt\, 
\delta \left({\hat c}_{i}(s)-{\hat c}_{j}(t)\right) 
\Big\} }
{ 
\int {\cal D}\hat{c} \,\,\, 
\exp\Big\{-\sum_{\alpha=0}^{n} H^{\alpha} + \frac{\mu ^{2}V}{2N} 
\sum_{i,j=1}^{N} \int_{0}^{1} \!\! ds\int_{0}^{1} \!\! dt\, 
\delta \left({\hat c}_{i}(s)-{\hat c}_{j}(t)\right) 
\Big\} 
}
=
\lim_{n \to 0}
\left\langle Q_{\hat{l}}
\right\rangle_{n+1}^{\rm P}, 
\label{EQ:Omega_Q}
\eeq
where $\langle\cdots\rangle_{n+1}^{\rm P}$ denotes an average for an
effective pure (\ie\ disorder-free) system of $n+1$ coupled replicas
of the original system, defined for a generic observable ${O}$ by
\begin{eqnarray}
\langle {O} \rangle_{n+1}^{\rm P} = 
\frac{
\int {\cal D}{\hat c} \,\,\,\, {O}
\,\,\,\, \exp\Big\{-\frac{1}{2}\sum_{j=1}^{N}\int_{0}^{1} \!\! ds
\left\vert \frac{d{\hat c}_{j}(s)}{{ds}}\right\vert^{2}
- N \tilde{\lambda}_{n}^{2} \frac{N}{V}
\tilde{\sum}^{\dagger}_{\hat{p}} \vert Q_{\hat p} \vert^{2} 
+ N \frac{\mu ^{2}}{V^{n}} {\overline{\sum}}^{\dagger}_{\hat{k}}  
\vert Q_{\hat p} \vert^{2} 
\Big\} 
} { 
\int {\cal D}{\hat c} \,\, 
\,\, \exp\Big\{-\frac{1}{2}\sum_{j=1}^{N}\int_{0}^{1} \!\! ds
\left\vert \frac{d{\hat c}_{j}(s)}{{ds}}\right\vert^{2}
- N \tilde{\lambda}_{n}^{2} \frac{N}{V}
\tilde{\sum}^{\dagger}_{\hat{p}} \vert Q_{\hat p} \vert^{2} 
+ N \frac{\mu ^{2}}{V^{n}} {\overline{\sum}}^{\dagger}_{\hat{k}}  
\vert Q_{\hat p} \vert^{2} 
\Big\} 
}.
\label{EQ:PureAv_def}
\end{eqnarray}

\subsection{Field theory}
\label{SEC:ft}
The expressions for the replicated partition function and the order
parameter presented in the last two subsections 
contain interactions between macromolecules
that complicate any analytic treatment. It therefore proves useful to 
decouple those
interactions by
performing a Hubbard-Stratonovich transformation that allows us to
eliminate the Fourier transformed density variables $Q_{\hat{k}}$ in
favor of field variables $\Omega_{\hat{k}}$~\cite{FNOTE:notation}. To
do that, we make use of the formulas:
\begin{eqnarray}
\exp \left( -a\left\vert{w}\right\vert^{2} \right)
&=&
\left(a/\pi\right)
\int d(\real z)\,\, d(\imag z) 
\,\,\exp \left( -a\left\vert{z}\right\vert^{2} \right)
\exp \left( 2ia\,\real zw^{\ast} \right),
\label{EQ:Hub_Strat_int}
\\
\exp \left( +a\left\vert{w}\right\vert^{2} \right)
&=&
\left(a/\pi\right) 
\int d(\real z)\,\,d(\imag z) 
\,\,\exp \left( -a\left\vert{z}\right\vert^{2} \right)
\exp \left( 2a\,\real zw^{\ast} \right)
\label{EQ:Hub_Strat_crs},
\end{eqnarray}
where $w$ is an arbitrary complex number, $a$ is a real and positive
(but otherwise arbitrary) number, and the integrals are taken over the
entire complex $z$ plane. 
We apply the lower formula to the
term originating in the crosslinking constraints, which comes with a positive
prefactor proportional to $\mu ^{2}$. 
We apply the upper formula to the
contributions with negative prefactors in the exponents in
Eq.~(\ref{EQ:Z_n_Q}), coming from
a combination of the excluded volume repulsions and the attractive 
effect of the constraints on the one replica sector.

By combining Eq.~(\ref{EQ:Z_n_Q}) with Eqs.~(\ref{EQ:Hub_Strat_int}) and
(\ref{EQ:Hub_Strat_crs}) in the manner just outlined, we obtain
\begin{equation}
[\bar{Z}^{n}] = {\cal N} \int\!\!{\cal D} \Omega \exp\{ -ndN{\cal
F}_{n}(\{\Omega_{\hat{k}}\}) \}, 
\label{EQ:Z_n_Omega}
\end{equation}
where ${\cal F}_{n}(\{\Omega_{\hat{k}}\})$ is a replicated free energy
functional. The symbol $\int\!\!{\cal D} \Omega$ denotes integration
over all possible configurations for the field $\Omega_{\hat{k}}$,
where the independent set of variables is the set of all
complex-valued $\Omega_{\hat{k}}$ (for $\hat{k}$ in the half-space
determined by the condition that $\hat{k} \cdot {\hat n}$ be positive
for a fixed unit vector ${\hat n}$). Outside of this half-space,
$\Omega_{\hat{k}}$ is defined by the relation $\Omega_{-\hat{k}} =
\Omega^{\ast}_{\hat{k}}$.  ${\cal N}$ is a normalization constant that
will be ignored henceforth, as it does not affect the value of the
order parameter [see Eq.~(\ref{EQ:op_av_Omega})] or the dependence of
the free energy on any shear deformations of the container (see
Sec.~\ref{SEC:f_Om_def}). As for the free energy functional ${\cal
F}_{n}(\{\Omega_{\hat{k}}\})$, it is given by
\begin{equation}
nd{\cal F}_{n}(\{\Omega_{\hat{k}}\}) 
= \tilde{\lambda}_{n}^{2} \frac{N}{V}
\tilde{\sum}^{\dagger}_{\hat{p}} \vert \Omega_{\hat p} \vert^{2} 
+ \frac{\mu ^{2}}{V^{n}} {\overline{\sum}}^{\dagger}_{\hat{k}}  
\vert \Omega_{\hat k} \vert^{2}
- \ln{ \left\langle 
\exp \left(   
i \tilde{\lambda}_{n}^{2} \frac{2 N}{V}
\tilde{\sum}^{\dagger}_{\hat{p}} \real \Omega_{\hat p} \rho^{\ast}_{\hat p}
+ \frac{2 \mu ^{2}}{V^{n}} {\overline{\sum}}^{\dagger}_{\hat{k}}  
\real \Omega_{\hat k} \rho^{\ast}_{\hat k}
\right) 
\right\rangle^{W}_{n+1}},
\label{EQ:F_Omega}
\end{equation}%
\eqresume%
where the one-macromolecule Fourier-transformed density $\rho_{\hat
k}$ is defined via
\begin{equation}
\rho_{\hat k} \equiv \int_{0}^{1} \!\!ds\,\, {\rm e}^{i {\hat k} \cdot 
{\hat c}(s) }
\label{EQ:rho_def}
\end{equation}
for a macromolecular configuration ${\hat c}(s)$, and the 
replicated Wiener average is defined by 
\begin{equation}
\left\langle O \right\rangle^{W}_{n+1} \equiv \frac{
\displaystyle
\int {\cal D}{\hat c} \,\, O 
\,\, \exp\Big(-\frac{1}{2} \int_{0}^{1} \!\! ds
\left\vert \frac{d{\hat c}(s)}{{ds}} \right\vert^{2}
\Big)
}{
\displaystyle
\int {\cal D}{\hat c} \,
\,\, \exp\Big(-\frac{1}{2} \int_{0}^{1} \!\! ds
\left\vert \frac{d{\hat c}(s)}{{ds}}\right\vert^{2}
\Big)
}.
\label{EQ:Wiener_def_n1}
\end{equation}

Analogously, the order parameter can be obtained as~\cite{FNOTE:notation}
\begin{equation}
\Omega_{{\bf k}^{1},\ldots,{\bf k}^{g}} = \lim_{n \to 0} \frac{ 
\int\!\!{\cal D} \Omega \,\,\, \Omega_{\hat{l}} \,\,\, 
\exp\{ -ndN{\cal F}_{n}(\{\Omega_{\hat{k}}\}) \} 
}{ 
\int\!\!{\cal D} \Omega 
\exp\{ -ndN{\cal F}_{n}(\{\Omega_{\hat{k}}\}) \}
},
\label{EQ:op_av_Omega}
\end{equation}
where, once again, we choose $\hat{l}$ so that ${\bf l}^{\alpha} = {\bf
k}^{\alpha}$ for $\alpha = 1, \ldots, g$ and ${\bf l}^{\alpha} = {\bf
0}$ for $\alpha = 0$ and $\alpha = g+1, \cdots, n$. 

\subsection{Stationary-point approximation}
The simplest available method to evaluate the free energy and the
order parameter is the stationary point approximation, which also
provides a starting point for possible improvement, for example by way
of the loop expansion~\cite{REF:Loop_Exp}. In the stationary point
approximation~\cite{FNOTE:MF_DG}
we have,
\begin{equation}
\bar{f}=d\lim_{n\rightarrow 0}\min_{\{\Omega_{\hat{k}}\}} 
{\cal F}_{n}\big(\{\Omega_{\hat{k}}\}\big).
\label{EQ:f_SP}
\end{equation}
The value $\bar{\Omega}_{\hat{l}}$ of the field ${\Omega}_{\hat{l}}$ that provides the minimum also
determines the order parameter according to:
\begin{equation}
\Omega_{{\bf k}^{1},\cdots,{\bf k}^{g}} = \lim_{n \to 0}
\bar{\Omega}_{\hat{l}}.
\label{EQ:op_SP}
\end{equation}
By demanding that the right hand side of Eq.~(\ref{EQ:F_Omega}) be stationary
with respect to variations of $\{\Omega_{\hat{k}}\}$, we obtain the
stationary point equations 
\begin{equation}
\bar{\Omega}_{\hat{p}} = i \Big\langle \rho_{\hat{p}}
\Big\rangle^{W,\bar{\Omega}}_{n+1}
\qquad \mbox{and} \qquad
\bar{\Omega}_{\hat{k}} = \Big\langle \rho_{\hat{k}}
\Big\rangle^{W,\bar{\Omega}}_{n+1}
\label{EQ:SPE} 
\end{equation}
for $\hat{p}$ lying in the one replica sector and $\hat{k}$ lying in
the higher replica sector. Here, we have defined the average
\begin{equation}
\langle O \rangle^{W,\bar{\Omega}}_{n+1} \equiv 
\frac{
\left\langle   O
\exp \left(   
i \tilde{\lambda}_{n}^{2} \frac{N}{V}
\tilde{\sum}_{\hat{p}} \bar{\Omega}_{\hat p} \rho^{\ast}_{\hat p}
+ \frac{\mu ^{2}}{V^{n}} {\overline{\sum}}_{\hat{k}}  
\bar{\Omega}_{\hat k} \rho^{\ast}_{\hat k}
\right)
\right\rangle^{W}_{n+1}
} {
\left\langle   
\exp \left(   
i \tilde{\lambda}_{n}^{2} \frac{N}{V}
\tilde{\sum}_{\hat{p}} \bar{\Omega}_{\hat p} \rho^{\ast}_{\hat p}
+ \frac{\mu ^{2}}{V^{n}} {\overline{\sum}}_{\hat{k}}  
\bar{\Omega}_{\hat k} \rho^{\ast}_{\hat k}
\right) 
\right\rangle^{W}_{n+1}
},
\label{EQ:WO_av_def} 
\end{equation}
\ie, an average over replicated macromolecular configurations under
the effect of a self consistency field $\bar{\Omega}_{\hat k}$.

\subsubsection{Proposing a general solution}
As has been discussed in Refs.~\cite{REF:epl,REF:cross}, the
stationary-point equation for the free-energy functional near the
transition is exactly solved by the following hypothesis:
\begin{mathletters}
\begin{eqnarray}
\Omega_{\hat{l}} & = &
(1-q)\,\delta_{ {\hat{l}},\hat{0}} 
+q\,\delta_{\tilde{\bf l},{\bf 0}}\,
\,\cpo^{\rm u}(\hat{l}),
\label{EQ:ord_par_hyp}
\\
\cpo^{\rm u}(\hat{l}) & \equiv & \int_{0}^{\infty}\!\!d\tau\,p(\tau)\,
{\rm e}^{-\hat{l}^{2}/2\tau}\,,
\label{EQ:cont_part_undef}
\end{eqnarray}
\end{mathletters}%
with
\begin{equation}
\tilde{\bf l} \equiv \sum_{\alpha=0}^{n}{\bf l}^{\alpha}.
\label{EQ:ktil_def}
\end{equation}%
The physical motivation for this hypothesis comes from a picture in
which a fraction $q$ (the ``gel fraction'') of the monomers are
localized around random mean positions ${\bf b}_{j}(s)$ about which
they execute harmonic thermal fluctuations characterised by random
localization lengths $\xi_{j}(s)$. In terms of the individual monomer
density of Eq.~(\ref{EQ:individual_density}), this picture translates
into the expression
\begin{equation}
\langle \exp(i {\bf k \cdot c}_{j}(s)) \rangle_{\chi} = \exp(i {\bf k
\cdot b}_{j}(s)) \exp(-{\xi_{j}(s)}^2 {\bf k}^2/2). 
\label{EQ:individual_density_gaussian}
\end{equation} 
The mean positions and localization lengths are assumed to be
distributed independently, with a homogeneous distribution over the
sample for the mean positions, and a statistical distribution
$2\xi^{-3} p(\xi^{-2})$ for the localization lengths. By combining the
contributions from all monomers in the system, we see that the
proposed statistical distributions give rise to an order parameter of
the form:
\eqbreak
\bml
\begin{eqnarray}
\Omega_{{\bf k}^{1},\cdots,{\bf k}^{g}} 
& = & (1-q) \,\, \delta_{{\bf k}^{1},{\bf 0}} \cdots \delta_{{\bf k}^{g},{\bf 0}} 
+ q \int \! \frac{d{\bf b}}{V} \,
{\rm e}^{ i{ ({\bf k}^{1} + \cdots + {\bf k}^{g}) \cdot {\bf b}} }
\int_{0}^{\infty}\!\!d({\xi^{-2}})\,p({\xi^{-2}})\, 
{\rm e}^{ -{\xi^{2}}( ({\bf k}^{1})^2 + \cdots + ({\bf k}^{g})^2 )/2}
\label{EQ:op_from_xi}
\\
& = & (1-q) \,\, \delta_{{\bf k}^{1},{\bf 0}} \cdots \delta_{{\bf
k}^{g},{\bf 0}}  
+ q \, 
\delta_{ {\bf k}^{1} + \cdots + {\bf k}^{g}, {\bf 0}}
\int_{0}^{\infty}\!\!d\tau\,p(\tau)\, 
{\rm e}^{-( ({\bf k}^{1})^2 + \cdots + ({\bf k}^{g})^2 )/2\tau}\,.
\label{EQ:op_from_dist}
\end{eqnarray}
\eml
The homogeneous distribution of the mean positions of the localized
particles gives rise, in Eq.~(\ref{EQ:op_from_xi}), 
to an integral over ${\bf b}$, 
which represents
the delta function that appears  
explicitly in the second term of the RHS of Eq.~(\ref{EQ:op_from_dist}). In the
second line we have also identified the variable $\tau = 1/\xi^{2}$.
By taking the replica limit in the manner of Eq.~(\ref{EQ:op_SP}), the
order parameter hypothesis [of Eqs.~(\ref{EQ:ord_par_hyp}) and
(\ref{EQ:cont_part_undef})] reduces to Eq.~(\ref{EQ:op_from_dist}).

{}From our motivation of the order parameter hypothesis, it is evident
that, in Eqs.~(\ref{EQ:op_from_xi}), (\ref{EQ:op_from_dist}), and
(\ref{EQ:ord_par_hyp}), delocalized and localized particles are,
respectively, represented by the first and second terms on the RHS.
The function $\cpo^{\rm u}(\hat{k})$,
which we refer to as the {\em continuous part\/} of the order parameter,
encodes information about thermal fluctuations (the
superscript u standing for ``unstrained system'').

The hypothesis of Eqs.~(\ref{EQ:ord_par_hyp}) and
(\ref{EQ:cont_part_undef}) for the order parameter only allows for a
liquid phase (for $q=0$) and for an amorphous solid phase (for $q >
0$).  It is useful to notice here that the order parameter is zero in
the one replica sector, independently of the values of $q$ and
$p(\tau)$.  This could have been anticipated, as in both the liquid and
the amorphous solid states, the order parameter is zero for values of
its argument lying in the one replica sector.

The hypothesis (\ref{EQ:ord_par_hyp}) and
(\ref{EQ:cont_part_undef}) solves the stationary-point
equations~(\ref{EQ:SPE}) if and only if the following self-consistent
equation for $q$ and $p(\tau)$ is satisfied:
\begin{mathletters}
\begin{eqnarray}
(\,1\,-\!&q&\!) \, \delta_{\hat{k},\hat{0}} 
+
q\,\delta_{\tilde{\bf k},{\bf 0}}
\,\,\int_{0}^{\infty}d\tau\,p(\tau)\,{\rm e}^{-\hat{k}^{2}/2\tau} 
\nonumber
\\
&=&
{\rm e}^{-\mu ^{2}q}\,
\delta_{\hat{k},\hat{0}}+
{\rm e}^{-\mu ^{2}q}\,
\delta_{\tilde{\bf k},{\bf 0}}
\int_{0}^{\infty} \!\!\! d\tau\,
{\rm e}^{-\hat{k}^{2}/2\tau}
\sum_{r=1}^{\infty}
{\mu ^{2r}q^{r}\over{r!}}
\int_{0}^{1} \!\! ds_{0}\cdots ds_{r}
\int_{0}^{\infty} \!\!\!
d\tau_{1}\cdots d\tau_{r}\,
p(\tau_{1})\cdots p(\tau_{r})\,
\delta\big(\tau-\Upsilon_{r}\big), 
\label{EQ:scone}
\\
\Upsilon_{r}^{-1}
& \equiv & 
W^{-1}
+S_{00}
-2W^{-1}\sum_{\rho=1}^{r}U_{\rho}\,S_{\rho 0}
-\sum_{\rho,\rho^{\prime}=1}^{r}
S_{0\rho}\,C_{\rho\rho^{\prime}}\,S_{\rho^{\prime}0},
\label{EQ:scmain}
\end{eqnarray}
\end{mathletters}
\eqresume
where
\bml
\begin{eqnarray} 
S_{\rho\rho^{\prime}} & \equiv & \min(s_{\rho},s_{\rho^{\prime}})
\qquad \mbox{(for $0\le\rho,\rho^{\prime}\le r$)}, \\
R_{\rho\rho^{\prime}} & \equiv &  
\tau_{\rho}^{-1}\delta_{\rho\rho^{\prime}}+S_{\rho\rho^{\prime}} 
\qquad \mbox{(for $1\le\rho,\rho^{\prime}\le r$)}, \\ 
U_{\rho} & \equiv &  \sum_{\rho^{\prime}=1}^{r}
R^{-1}\vert_{\rho\rho^{\prime}},  \\
W & \equiv & \sum_{\rho,\rho^{\prime}=1}^{r}
R^{-1}\vert_{\rho\rho^{\prime}},\mbox{ and } \\
C_{\rho\rho^{\prime}} & \equiv  & 
R^{-1}\vert_{\rho\rho^{\prime}}-W^{-1}U_{\rho}\,U_{\rho^{\prime}}.
\end{eqnarray} 
\eml
By passing to the limit $\hat{k}^{2}\rightarrow 0$, via a sequence for
which $\tilde{\bf k}={\bf 0}$, we identify from Eq.~(\ref{EQ:scone}) 
the self-consistency condition for
the gel-fraction $q$~\cite{REF:Erdos}:
\begin{equation}
q=1-{\rm e}^{-\mu ^{2}q}. 
\label{EQ:scforQ}
\end{equation}
For $\mu ^{2}\le 1$ (\ie\ for sufficiently small densities of
crosslinks), Eq.~(\ref{EQ:scforQ}) has only the trivial solution
$q=0$, corresponding to the liquid state. However, for $\mu ^{2}>1$
there are two solutions: one is still $q=0$; the other is
$q=\bar{q}(\mu ^{2}) > 0$, corresponding to the amorphous solid
state. The solution $\bar{q}(\mu ^{2})$ continuously bifurcates from
the liquid solution at $\mu ^{2}=1$, and approaches unity,
asymptotically, for large $\mu ^{2}$.
For $\mu ^2 >1$, it is straightforward to show, by expanding the free energy
functional ${\cal F}(\{\Omega_{\hat{k}}\})$ of Eq.~(\ref{EQ:F_Omega})
to second order in the variables $\{\Omega_{\hat{k}}\}$, that the
liquid state stationary point $\bar{\Omega}^{\rm liq}_{\hat{k}} \equiv
\delta_{\hat{k},\hat{0}}$ is unstable, and the equilibrium state is
therefore certainly not a liquid.

The fact that $q=\bar{q}(\mu ^{2})$ in the amorphous solid state grows
continuously from zero at the transition, together with the form of
$\Omega_{\hat{k}}$ in Eq.~(\ref{EQ:ord_par_hyp}) tells us that the
transition is continuous. Both the gel fraction and the order
parameter (for nonzero wavevectors) can be taken as small quantities in
the regime close to the transition. In what follows, we exploit this
information to analyze this regime in more detail.

\subsection{Behavior near the amorphous solidification transition}
For the regime close to the transition which occurs at $\mu ^2 = 1$, it is
convenient to define a variable $\epsilon \equiv 3(\mu ^2-1)$ that
measures the distance to the transition. For $\epsilon < 0$ the system
is in the liquid phase, and for  $\epsilon > 0$ the system is in the
amorphous solid phase. 

To obtain more detailed information about the properties of the system
close to the transition, two equivalent approaches are available at this
point.  One is to simplify the self-consistent equation
(\ref{EQ:scone}) for $q$ and $p(\tau)$ by using the knowledge that
$q$ is small near the transition to neglect all powers $r > 2$ in
the right hand side. The second one is to go back to the expression
for the free energy functional, Eq.~(\ref{EQ:F_Omega}), expand in
powers of $\{\Omega_{\hat{k}}\}$, and neglect all powers 
higher than the third. In Refs.~\cite{REF:epl,REF:cross}, the former
approach was taken, but here we are going to use the later approach,
because it allows for a more straightforward determination of the
effects of deformations on the system.

\subsubsection{Free energy functional}
\label{SEC:exp_F}
As that saddle point value of the order parameter is zero for
wavevectors lying in the one replica sector, any term in the expansion
of ${\cal F}_{n}\big(\{\Omega_{\hat{k}}\}\big)$ that contains it as a
factor will automatically vanish. We therefore ignore all such terms
henceforth.

Close to the transition the order parameter is dominated by long
localization lengths (we shall show later that the typical localization
length diverges at the transition). This is to be expected on physical
grounds, because the system is ``barely solid'', allows the monomers
to thermally fluctuate over long distances; and it has also been shown
directly by solving for $p(\tau)$ in Eq.~(\ref{EQ:scone})
\cite{REF:epl,REF:cross}.  Here we take this as an assumption, and
later show that the solution obtained for the order parameter is
consistent with this assumption~\cite{FNOTE:other_solutions}.

Thus, by expanding Eq.~(\ref{EQ:F_Omega}) in powers of the order
parameter and the wave vectors, assuming that the order parameter is
zero in the one replica sector (and, in order to simplify later
algebra, rescaling $\cal F$ by an overall factor of $6$), we obtain
for the regime near the transition a free energy of the form:
\begin{eqnarray}
\lefteqn{
nd{\cal F}_{n}\big(\{\Omega_{\hat{k}}\}\big) = 
{\overline{\sum}}_{\hat{k} \in  R^{\rm u} }
\big(-\epsilon+\frac{1}{2}|\hat{k}|^2\big)
\big\vert\Omega_{\hat{k}}\big\vert^{2}
}
\nonumber\\
&&\qquad\quad
-\,{\overline{\sum}}_{{\hat{k}_1}{\hat{k}_2}{\hat{k}_3} \in  R^{\rm u} }
\Omega_{\hat{k}_1}\,
\Omega_{\hat{k}_2}\,
\Omega_{\hat{k}_3}\,
\delta_{{\hat{k}_1}+{\hat{k}_2}+{\hat{k}_3}, {\hat{0}}}\,.
\label{EQ:free_en_unstrained}
\end{eqnarray}
This form for ${\cal F}_{n}(\{\Omega_{\hat{k}}\})$ can be obtained
either from a semi-microscopic model, as sketched here, or via an
argument involving symmetries and the continuity of the transition in
the context of a Landau theory. The same free energy functional
actually describes a universality class of physical systems that
display liquid--amorphous-solid transitions similar to the one shown by
vulcanized systems~\cite{REF:landau}.

\subsubsection{Stationary-point approximation}
We now obtain the stationary point approximation by demanding that
variations of ${\cal F}_{n}(\{\Omega_{\hat{k}}\})$ with respect to the
order parameter should be zero. This results in the stationarity
equation: 
\begin{equation}
 0 = 
 2\big(-\epsilon+\frac{1}{2}|{\hat{k}}|^2\big)
 \Omega_{\hat{k}} 
  -3{{\displaystyle{\overline\sum}}
	\atop{\scriptstyle\hat{k}_1\hat{k}_2\in R^{\rm u} }}
 \Omega_{\hat{k}_1}\,
 \Omega_{\hat{k}_2}\,
 \delta_{{\hat{k}_1}+{\hat{k}_2},{\hat{k}}}\,.
 \label{EQ:LG_saddle_unstrained}
\end{equation}

The stationary point equation Eq.~(\ref{EQ:LG_saddle_unstrained}) is 
satisfied (in the limit $n\rightarrow 0$) by the hypothesis
Eqs.~(\ref{EQ:ord_par_hyp}) and (\ref{EQ:cont_part_undef}), provided
that
\begin{eqnarray}
&&0=\delta_{\tilde{\bf k},{\bf 0}}
\left\{ 
2\left(3q^{2}-\epsilon q+q
{{\hat{k}}^2}/{2}
\right)
\int_{0}^{\infty}\!\!d\tau\,p(\tau)\,
\,{\rm e}^{-\hat{k}^{2}/2\tau}
\right. 
\nonumber\\
&&\,\,-\left.
3q^{2}\!
\int_{0}^{\infty}\!\!\!d\tau_1\,p(\tau_1) 
\int_{0}^{\infty}\!\!\!d\tau_2\,p(\tau_2)
\,{\rm e}^{-\hat{k}^{2}/2(\tau_1+\tau_2)}\right\}.
\nonumber\\
\label{EQ:saddle_plug}
\end{eqnarray}

\subsubsection{Gel fraction}
By taking the limit ${\hat{k}}^{2}\rightarrow 0$, the above equation
reduces to a condition for the gel fraction $q$, 
\begin{equation}
0=-2q\epsilon+3q^2. 
\label{EQ:q_eqn}
\end{equation}
For negative or zero $\epsilon$, corresponding to a crosslink density
less than or equal to its critical value, the only physical solution
(\ie\ with $0 \le q \le 1$) is
$q = 0$, corresponding to the liquid state. 
For positive $\epsilon$, corresponding to a crosslink
density in excess of the critical value, there are two solutions. One,
unstable, is the continuation of the liquid state $q=0$; the other,
stable, corresponds to a nonzero gel fraction, \ie\ to the amorphous
solid state, 
\begin{equation}
q=\frac{2}{3}\epsilon.
\label{EQ:q_soln}
\end{equation}
As mentioned above, the gel fraction (and consequently the order
parameter) change continuously at the transition, which means that at
$\epsilon=0$ there is a continuous phase transition between the liquid
and the amorphous solid state. Moreover, the linear dependence of the
gel fraction with $\epsilon$ implies a similar linear dependence with
the excess of the crosslink density $[M]/N$ above the critical value $M_{\rm
c}/N$ at the transition: $q \approx 2\epsilon/3 \sim \big([M]/M_{\rm
c}-1\big) = \big([M]/M_{\rm c}-1\big)^{\beta} $, with $\beta=1$,
\ie\ we recover the classical exponent for the gel fraction.

\subsubsection{Distribution of localization lengths}
\label{SEC:pi_theta}
In the amorphous solid state, by assuming that
Eq.~(\ref{EQ:q_eqn}) is satisfied, Eq.~(\ref{EQ:saddle_plug}) reduces
to a nonlinear integro-differential equation involving only the
distribution of (inverse square) localization lengths $p(\tau)$:
\begin{equation}
\frac{\tau^2}{2}\frac{dp}{d\tau}=
\Big(\frac{\epsilon}{2}-\tau\Big)p(\tau)
-\frac{\epsilon}{2}\int_{0}^{\tau}\!\!d\tau_1\,
 p(\tau_1)\,p(\tau-\tau_1).
\label{EQ:scp}   
\end{equation}
The form of this equation immediately suggests that, to the present
level of approximation, all $\epsilon$ dependence can be eliminated by
the scalings~\cite{FNOTE:higher_order}:
\begin{equation}
p(\tau)=({2}/{\epsilon})\,\pi(\theta);
\qquad
\tau=({\epsilon}/{2})\,\theta.
\label{EQ:p_scale}
\end{equation} 
Thus, the universal scaling function $\pi(\theta)$ satisfies the
parameter free equation
\begin{equation}
\frac{\theta^{2}}{2} \frac{d\pi}{d\theta}
= (1-\theta)\,\pi(\theta)-
\int_{0}^{\theta} d\theta^{\prime}
\pi(\theta^{\prime})\pi(\theta-\theta^{\prime}), 
\label{EQ:scpieq}
\end{equation}
together with the normalization condition 
\begin{equation}
1=\int\nolimits_{0}^{\infty}d\theta\,\pi(\theta).
\label{EQ:pi_norm}
\end{equation}
This normalization condition directly follows from the fact
that the order parameter of Eq.~(\ref{EQ:opDefinition}) has to be
unity at the origin~\cite{REF:epl,REF:cross}, and is consistent with
the physical interpretation of $p(\tau)$ as a probability
distribution. 

The scaling function $\pi(\theta)$ determines the behavior of both the
distribution of localization lengths and the order parameter near the
transition.  It has a peak at $\theta\simeq 1$ of width of order
unity, and decays rapidly both as $\theta\to 0$ and $\theta\to\infty$.
The asymptotic forms of the decays are: $\pi(\theta)\sim
a\theta^{-2}\exp\big(-2/\theta\big)$ (for $\theta\ll 1$) and
$\pi(\theta)\sim 3\big(b\theta-3/5\big)\exp\big(-b\theta\big)$ (for
$\theta\gg 1$). These forms are obtained analytically from
Eq.~(\ref{EQ:scpieq}); the coefficients $a\approx 4.554$ and $b\approx
1.678$ can be extracted from the complete numerical solution of
Eq.~(\ref{EQ:scpieq})~\cite{REF:epl}.

Due to  the fact that $\pi(\theta)$ has a well-defined, unique peak that
concentrates most of the weight, it makes sense to define a {\em
typical\/} localization length  $\xi_{\rm typ}$, and from
Eq.~(\ref{EQ:p_scale}) we see that it scales as $\xi_{\rm typ} \sim 
\epsilon^{-1/2}$. In particular, it diverges at the
transition, as anticipated above.  It is interesting to notice that
this typical length scales with the same exponent as the one obtained
in the classical theory for the correlation length. However, $\xi_{\rm
typ}$ is a quantity that describes the localized
monomers, whereas the correlation length of the classical theory
describes the delocalized monomers.

\subsubsection{Order parameter}
The order parameter also has a scaling form near the transition, which
follows directly from its parametrization in terms of $q$ and
$p(\tau)$, Eqs.~(\ref{EQ:ord_par_hyp}) and (\ref{EQ:cont_part_undef}),
and the scaling form for $p(\tau)$, Eq.~(\ref{EQ:p_scale}):
\bml
\begin{eqnarray}
\Omega_{\hat{k}}&=&
\big(1-2\epsilon/3\big)\,\delta_{\hat{k},\hat{0}} \!+\! 
\big(2\epsilon/3\big)\,\delta_{\tilde{\bf k},{\bf 0}}\,
\omega
\!\Big(\sqrt{2\hat{k}^{2}/\epsilon}\Big),
\label{EQ:omsc}\\
\omega(k)&=&
\int_{0}^{\infty} \!\!d\theta\,\pi(\theta)\,
{\rm e}^{-k^{2}/2\theta}.
\label{EQ:omdef}
\end{eqnarray}
\eml
Hence, we see that the  order parameter is  also  described in terms of  a
scaling function, in this case $\omega(k)$. As  for 
$\pi(\theta)$, the  asymptotic forms of $\omega(k)$ can be obtained
analytically: $\omega(k)\sim 1+c k^{2}+d k^{4}$ (for $k\ll 1$) and 
$\omega(k)\sim \big(9\pi k^{3}/\sqrt{8b}\big)^{1/2}
\exp\big(-\sqrt{2bk^{2}}\big) \big\{1+(27/40\sqrt{2bk^{2}})\big\}$ 
(for $k\gg 1$). A numerical calculation yields $\omega(k)$ for all
$k$, and determines the coefficients $c \approx
-0.4409$ and $d \approx 0.1316$ (see Ref.~\cite{REF:epl}). 

\section{Response to shear strain} 
\label{SEC:response}
In this section we discuss the effects of an externally applied
strain, both on the semi-microscopic macromolecular structure of the
system and on the value of its free energy.
To do this, we repeat the procedure followed in the previous section
to obtain the order parameter and the free energy of the system, but
this time we consider in detail the effects of deforming the
boundaries of the container. 
As we did before, we are going to concentrate on the behavior near the
amorphous solidification transition, and we are going to employ the
stationary point approximation in order to obtain explicit results.

\subsection{Description of the deformation}

We characterize the deformation of the system by the ($d\times d$)
deformation 
matrix ${\fft S}$, that describes the change in position of any point
${\bf b}$ at the boundary of the system ${\bf b}\rightarrow{\fft
S}\cdot {\bf b}$, with ${\fft S}$ independent of ${\bf b}$.  
For any matrix ${\fft S}$, it is possible to find a diagonal
matrix ${\fft \bar{S}}$ and two rotations ${\fft U}$ and ${\fft V}$
such that the decomposition ${\fft S} = {\fft U} {\fft\bar{S}} {\fft
V}$ holds~\cite{FNOTE:decompose}. 
This decomposition can be interpreted in terms of a physical
process performed in three steps: in the first, the system is rotated
in space as described by ${\fft V}$; in the second, it is deformed with
the diagonal deformation matrix ${\fft \bar{S}}$; and in the third, it
is rotated as 
described by ${\fft U}$. The only part of this process that represents
a genuine strain, and can therefore possibly alter the free energy of the 
system, is the second step. Therefore, we may (and shall) always assume, 
without loss of generality, that the deformation matrix is diagonal.
As an example of a deformation matrix for $d=3$, let us consider the
case in which the $x$, $y$ and
$z$ Cartesian components of the position vector are, respectively,
elongated by the factors $\lambda_{x}$, $\lambda_{y}$ and
$\lambda_{z}$, the matrix ${\fft S}$ has the form ${\rm
diag}(\lambda_{x},\lambda_{y},\lambda_{z})$.  As we are concerned with
the effects of pure shear strains, we shall consider only deformations
that leave the volume $V$ of the system unchanged, \ie,
\begin{equation}
{\rm det}\,{\fft S}=1.
\label{EQ:det_one}
\end{equation}
For considering infinitesimal strains, it is convenient to define the
(diagonal) strain tensor
\begin{equation}
{\fft J}\equiv{\fft S}-{\fft I},
\label{EQ:small_def}
\end{equation}
where ${\fft I}$
is the identity matrix. For small shear deformations, we have 
\begin{equation}
1 = {\rm det}\,{\fft S} = 1 + {\rm tr}\,({\fft S} - {\fft I}) + {\cal
O}(({\fft S} - {\fft I})^{2}),  
\label{EQ:detS_expand}
\end{equation}
and consequently
\begin{equation}
{\rm tr}\,{\fft J}=0, 
\label{EQ:no_trace}
\end{equation}
to first order in the deformation.

\subsection{Deformation and replicas}
Before taking the thermodynamic limit, the system is finite in extent,
and thus the Fourier representation of any function of position
consists of a superposition of plane waves with wave-vectors belonging
to a discrete set.  The precise set of wave vectors is determined by the
periodic boundary conditions.  In particular, the order parameter is
represented by a function $\Omega_{\hat{k}}$ that is only defined at a
discrete set of points in replicated Fourier space.  Now, under strain
the boundaries in position space are displaced and, as a consequence,
the discrete set of points in replicated Fourier space move.  As previously
mentioned in Sec.~\ref{SEC:av_rep}, any external strain applied to the
system after the permanent constraints have been created will affect
replicas $\alpha=1,\ldots,n$, but not replica
$\alpha=0$~\cite{REF:DeamEd}.  Therefore, the change in the
discretization of the wave vectors occurs only for
$\alpha=1,\ldots,n$, but not $\alpha=0$. For replicas
$\alpha=1,\ldots,n$, the set of allowed $d$-dimensional wave vectors
$r^{\rm u}$ corresponding to the unstrained system is replaced by a
new set $r^{\rm s}$ corresponding to the strained system.
Consequently, the set $R^{\rm s}$ of allowed {\it replicated\/} wave-vectors
in the strained system is composed of all replicated wave vectors
${\hat{k}} = \{{\bf k}^{0},{\bf k}^{1},\cdots,{\bf k}^{n}\}$ such that
${\bf k}^{0} \in r^{\rm u}$ and ${\bf k}^{\alpha} \in r^{\rm s}$ (for
replicas $\alpha=1,\ldots,n$).

\subsection{Free energy functional for the deformed system}
\label{SEC:f_Om_def}
Conceptually, there are two sources for the change in free energy,
Eq.~(\ref{EQ:f_SP}), under deformation: the change in the
expression for the free energy functional itself, and the consequent
change in the value of the order parameter that solves the
stationary-point equation.  The free-energy functional for the strained
system ${\cal F}_{n}^{\rm s}(\{\Omega_{\hat{k}}\})$ is obtained by
repeating, step-by-step, the procedure followed in Secs.~\ref{SEC:av_rep} and
\ref{SEC:ft} to construct the free-energy
functional for the unstrained system ${\cal
F}_{n}(\{\Omega_{\hat{k}}\})$.  The only change resides in the fact that
integrals over the positions of the monomers in replicas $\alpha = 1,
\ldots, n$ now range over the region
occupied by the strained sample instead of the region occupied by the
unstrained sample and, consequently, the periodic delta function of
Eq.~(\ref{EQ:delta_rep}) now involves a 
summation over the new set $R^{\rm s}$ of wave vectors in replicated space:
\begin{equation}
\delsh({\hat c}) = \frac{1}{V^{1+n}} \sum_{{\hat p} \in R^{\rm s}}
\exp\left(i{\hat p}\cdot {\hat c}\right).
\label{EQ:delta_rep_def}
\end{equation}
As a result, Eq.~(\ref{EQ:int_Qh}) is replaced by 
\begin{equation} 
\frac{\mu ^{2} V}{2 N} \sum_{i,j=1}^{N} \int_{0}^{1} \!\! ds\int_{0}^{1}
\!\! dt\, \delsh \left({\hat c}_{i}(s)-{\hat c}_{j}(t)\right) 
= 
\frac{\mu ^{2} N}{2 V^{n}} \sum_{{\hat p} \in R^{\rm s}} \vert Q_{\hat
p} \vert^{2}, 
\label{EQ:int_Qh_def}
\end{equation}
and the expression for $[\bar{Z}^{n}]$ in terms of monomer densities
given in Eq.~(\ref{EQ:Z_n_Q}) is replaced by
\eqbreak
\begin{equation}
[\bar{Z}^{n}] 
= 
\frac{
{\rm e}^{-Nn \phi}
\int {\cal D}{\hat c}
\,\,\,\exp\Big\{-\frac{1}{2}\sum_{j=1}^{N}\int_{0}^{1} \!\! ds
\left\vert \frac{d{\hat c}_{j}(s)}{{ds}}\right\vert^{2}
- N \tilde{\lambda}_{n}^{2} \frac{N}{V}
\tilde{\sum}^{\dagger}_{{\hat p} \in R^{\rm s}} \vert Q_{\hat p} \vert^{2} 
+ N \frac{\mu ^{2}}{V^{n}} {\overline{\sum}}^{\dagger}_{\hat{k} \in R^{\rm s}}  
\vert Q_{\hat k} \vert^{2} 
\Big\} }
{ 
\int {\cal D}{\bf c} \,\,\,
\exp\big\{-\frac{1}{2}\sum_{j=1}^{N}\int_{0}^{1} \!\! ds 
\left\vert \frac{d{\bf c}_{j}(s)}{{ds}}\right\vert^{2} -
N \tilde{\lambda}_{0}^{2} \frac{N}{V}
\sum^{\dagger}_{\bf p} \vert Q_{\bf p} \vert^{2}
\big\} 
}.
\label{EQ:Z_n_Q_def}
\end{equation}
Two features should be noted here. One is that the denominator in
formula (\ref{EQ:Z_n_Q_def}) is not affected by the deformation,
because it is
the normalization factor for the disorder distribution, which is fixed
before the system is deformed. Thus, the normalization constant $\cal
N$ in Eq.~(\ref{EQ:Z_n_Omega}), which reads
\begin{equation}
{\cal N} = \frac{ \exp(-Nn\phi) }{ \displaystyle \int {\cal D}{\bf c} \,\,\,
\exp
Big\{
-\frac{1}{2}\sum_{j=1}^{N}\int_{0}^{1} \!\! ds 
\left\vert \frac{d{\bf c}_{j}(s)}{{ds}}\right\vert^{2} -
N \tilde{\lambda}_{0}^{2} \frac{N}{V}
\sum^{\dagger}_{\bf p} \vert Q_{\bf p} \vert^{2}
\Big\}},
\label{EQ:cal_N}
\end{equation}
is unchanged by the deformation, as anticipated in Sec.~\ref{SEC:ft}.
The second feature is that no changes have appeared in any of the
prefactors in front of the terms in the exponent in
the numerator of Eq.~(\ref{EQ:Z_n_Q}) that {\em
are\/} affected by the deformation.

{}From Eq.~(\ref{EQ:Z_n_Q_def}) one immediately obtains, with the 
Hubbard-Stratonovich transformation as in Sec.~\ref{SEC:ft}, the free
energy functional 
\begin{equation}
nd{\cal F}_{n}^{\rm s}(\{\Omega_{\hat{k}}\}) 
= \tilde{\lambda}_{n}^{2} \frac{N}{V}
\tilde{\sum}^{\dagger}_{\hat{p} \in R^{\rm s}} \vert \Omega_{\hat p} \vert^{2} 
+\! \frac{\mu ^{2}}{V^{n}} {\overline{\sum}}^{\dagger}_{\hat{k} \in R^{\rm s}}  
\vert \Omega_{\hat p} \vert^{2}
-\! \ln{ \!\left\langle 
\exp \!\left(   
i \tilde{\lambda}_{n}^{2} \frac{2 N}{V}
\tilde{\sum}^{\dagger}_{\hat{p} \in R^{\rm s}} \real \Omega_{\hat p}
\rho^{\ast}_{\hat p} 
+\! \frac{2 \mu ^{2}}{V^{n}} {\overline{\sum}}^{\dagger}_{\hat{k} \in R^{\rm s}}  
\real \Omega_{\hat k} \rho^{\ast}_{\hat k}
\right) 
\right\rangle^{W}_{n+1}}.
\label{EQ:F_Omega_def}
\end{equation}
\eqresume
As in the case of the undeformed system, we are going to
take one further step, and restrict ourselves 
to the regime near the amorphous solidification transition.

\subsection{Free energy and stationary point equations near the
vulcanization transition} 
In the regime close to the transition, we can expand the free energy
functional in powers of the
order parameter and the wave vectors, as we did in
Sec.~\ref{SEC:exp_F}, and obtain the analog of
Eq.~(\ref{EQ:free_en_unstrained}) for the deformed system:
\begin{eqnarray}
\lefteqn{
nd{\cal F}_{n}^{\rm s}\big(\{\Omega_{\hat{k}}\}\big) = 
{\overline{\sum}}_{\hat{k} \in  R^{\rm s} }
\big(-\epsilon+\frac{1}{2}|\hat{k}|^2\big)
\big\vert\Omega_{\hat{k}}\big\vert^{2}
}
\nonumber\\
&&\qquad\qquad
-\,{\overline{\sum}}_{{\hat{k}_1}{\hat{k}_2}{\hat{k}_3} \in  R^{\rm s} }
\Omega_{\hat{k}_1}\,
\Omega_{\hat{k}_2}\,
\Omega_{\hat{k}_3}\,
\delta_{{\hat{k}_1}+{\hat{k}_2}+{\hat{k}_3}, {\hat{0}}}\,.
\label{EQ:free_en_def}
\end{eqnarray}

As a result, the stationary-point
equation for the strained system becomes
\begin{equation}
 0 = 
 2\Big(-\epsilon+\frac{1}{2}|{\hat{k}}|^2\Big)
 \Omega_{\hat{k}} 
  -3{{\displaystyle{\overline\sum}}
	\atop{\scriptstyle\hat{k}_1\hat{k}_2\in R^{\rm s} }}
 \Omega_{\hat{k}_1}\,
 \Omega_{\hat{k}_2}\,
 \delta_{{\hat{k}_1}+{\hat{k}_2},{\hat{k}}}\,.
 \label{EQ:LG_saddle}
\end{equation}
Although, superficially, this equation looks the same as
Eq.~(\ref{EQ:LG_saddle_unstrained}), they are actually different,
as all the wavevectors entering in
Eq.~(\ref{EQ:LG_saddle_unstrained}) belong to $R^{\rm u}$, \ie\ the set
of replicated wavevectors corresponding to the undeformed system, and
all the wavevectors entering in Eq.~(\ref{EQ:LG_saddle}) belong to
$R^{\rm s}$, \ie\ the set of replicated wavevectors corresponding to
the deformed system. Therefore, whilst 
Eq.~(\ref{EQ:LG_saddle_unstrained}) is invariant under all
permutations of the $1+n$ replicas, Eq.~(\ref{EQ:LG_saddle}) is only
invariant under permutations of the $n$ replicas $\alpha =1, \ldots,
n$.

\subsection{Proposing a hypothesis for the order parameter}
We shall obtain the order parameter 
for the strained state
by finding a solution of 
Eq.~(\ref{EQ:LG_saddle}).  To do this, we shall use physical arguments
similar to those used in the case of the unstrained system to motivate
our guess for a possible solution. As our guess will turn out to solve
Eq.~(\ref{EQ:LG_saddle}) exactly, this justifies, {\em a posteriori\/},
our physical assumptions. As the shear modulus is determined by an
expansion of the free energy to quadratic order in the deformation,
for the moment we will only consider infinitesimal deformations.

\begin{figure}[hbt]
 \epsfxsize=3.0in
 \centerline{\epsfbox{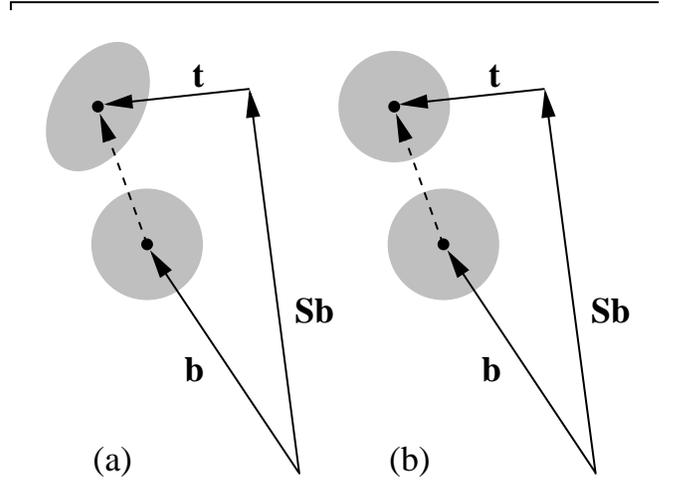}} 
\vspace{0.5cm}
\caption{Change of the localization region for an individual monomer
due to an externally imposed strain: (a) assuming that the fluctuation
region around the mean position deforms affinely; (b) assuming that the
fluctuation region around the mean position stays spherically symmetric.
}
\label{FIG:defo_hypo}
\end{figure}%
For each localized monomer in the unstrained system we envisage that
its old mean position ${\bf b}_{j}(s)$ is displaced to a new mean
position ${\bf b}^{\rm s}_{j}(s) = {\fft S}\cdot {\bf b}_{j}(s) + {\bf
t}_{j}(s)$. Up to this point the only assumption is the physically
intuitive one that those monomers that are localized in the undeformed
system remain localized in the deformed system. The vector ${\fft
S}\cdot {\bf b}_{j}(s)$ is the affine displacement of the old
position~\cite{REF:Florybook}. We now make the assumption that 
${\bf t}_{j}(s)$ is a random additional displacement, uncorrelated with
${\bf b}_{j}(s)$.

For each localized monomer, we also need some conjecture about 
the behavior under strain of 
the size and shape of the region within which it thermally fluctuates. We
assume that this localization region need not remain spherical (as it
was in the unstrained system) but might be deformed due to the
external strain. We will consider the position fluctuations for the
monomers:
\begin{equation}
\delta {\bf c}^{\rm u}_{j}(s) = {\bf c}_{j}(s) - {\bf b}_{j}(s) 
\label{EQ:r_fluct_u}
\end{equation}
for the unstrained system, and
\begin{equation}
\delta {\bf c}^{\rm s}_{j}(s) = {\bf c}_{j}(s) - ({\fft S}\cdot {\bf
b}_{j}(s) + {\bf t}_{j}(s)) 
\label{EQ:r_fluct_s}
\end{equation}
for the strained system, and also the individual monomer densities for
the unstrained and strained systems, $\rho^{\rm u}_{j,s,\chi}({\bf r})$
and $\rho^{\rm s}_{j,s,\chi}({\bf r})$, as defined by
Eq.~(\ref{EQ:density_r_def}).

One possible assumption is that the fluctuation region deforms
affinely, \ie\ that
\begin{equation}
\delta {\bf c}^{\rm u}_{j}(s) \rightarrow \delta {\bf c}^{\rm
s}_{j}(s) = {\fft S} \cdot \delta {\bf c}^{\rm u}_{j}(s). 
\label{EQ:affine_cloud}
\end{equation}
This gives rise to the individual monomer density
\begin{equation}
\rho^{\rm s}_{j,s,\chi}({\bf r}) = \rho^{\rm u}_{j,s,\chi}({\fft S}^{-1}({\bf
r} - {\bf b}^{\rm s}_{j}(s)) + {\bf b}_{j}(s)) 
\label{EQ:affine_den_r}
\end{equation}
in real space, and
\begin{eqnarray}
\langle {\rm e}^{i{\bf k}\cdot {\bf c}_{j}(s)}\rangle^{\rm s}_{\chi} & = &
\exp\big( i{\bf k}\!\cdot \! \left\{ {\fft S}\!\cdot \!{\bf b}_{j}(s)
+ {\bf t}_{j}(s) \right\} \big)
\nonumber \\
&& 
\times \exp\big(-\xi^{2}_{j}(s)\,{{\bf k}\!\cdot \! 
{\fft S}^{2} 
\!\cdot \!{\bf k}}/2\big)
\label{EQ:affine_den_k}
\end{eqnarray}
for the Fourier-transformed version.  In what follows, we will replace
the matrix $ {\fft S}^{\rm 2}$ by its expansion to first order in the
deformation
\begin{equation}
{\fft S}^{2} \approx {\fft I} + 2 {\fft J} 
+ {\cal O}({\fft J}^{2}). 
\label{EQ:STS_expand}
\end{equation}
Thus, for an infinitesimal strain, the assumption of affine distortion
of the fluctuation region gives the density 
\begin{eqnarray}
\langle {\rm e}^{i{\bf k}\cdot {\bf c}_{j}(s)}\rangle^{\rm s}_{\chi} & = &
\exp\big( i{\bf k}\!\cdot \! \left\{ {\fft S}\!\cdot \!{\bf b}_{j}(s)
+ {\bf t}_{j}(s) \right\} \big)
\nonumber \\
&& 
\times \exp\big(-\xi^{2}_{j}(s)\,{{\bf k}\!\cdot \! \left\{ 
{\fft I} + 2 {\fft J} 
\right\} \!\cdot \!{\bf k}}/2\big).
\label{EQ:affine_den_k_linear}
\end{eqnarray}
 
An alternative assumption is that the fluctuation region remains
spherical as in the unstrained system, \ie, that 
\begin{equation}
\delta {\bf c}^{\rm u}_{j}(s) \rightarrow \delta {\bf c}^{\rm
s}_{j}(s) = \delta {\bf c}^{\rm u}_{j}(s).
\label{EQ:spherical_cloud}
\end{equation}
This, in turn, gives rise to the individual monomer density
\begin{equation}
\rho^{\rm s}_{j,s,\chi}({\bf r}) = \rho^{\rm u}_{j,s,\chi}({\bf
r} - {\bf b}^{\rm s}_{j}(s) + {\bf b}_{j}(s) )
\label{EQ:spherical_den_r}
\end{equation}
in real space, and
\begin{eqnarray}
\langle {\rm e}^{i{\bf k}\cdot {\bf c}_{j}(s)}\rangle^{\rm s}_{\chi} & = &
\exp\big( i{\bf k}\!\cdot \! \left\{ {\fft S}\!\cdot \!{\bf b}_{j}(s)
+ {\bf t}_{j}(s) \right\} \big)
\nonumber \\
&& 
\times \exp\big(-\xi^{2}_{j}(s) {{\bf k}^{2}}/2\big)
\label{EQ:spherical_den_k}
\end{eqnarray}
in wave vector space.

Motivated by the above special cases, we propose the following
parametrization for $\langle {\rm e}^{i{\bf k}\cdot {\bf
c}_{j}(s)}\rangle^{\rm s}_{\chi}$, which contains 
Eqs.~(\ref{EQ:affine_den_k_linear}) and (\ref{EQ:spherical_den_k}) as
particular limits: 
\begin{eqnarray}
\langle {\rm e}^{i{\bf k}\cdot {\bf c}_{j}(s)}\rangle^{\rm s}_{\chi} 
& = &
\exp\big( i{\bf k}\!\cdot \! \left\{ {\fft S}\!\cdot \!{\bf b}_{j}(s)
+ {\bf t}_{j}(s) \right\} \big)
\nonumber \\
&& 
\times \exp\big(-\xi^{2}_{j}(s)\,{{\bf k}\!\cdot \! \left\{ {\fft I}
		+\eta_{j}(s) \, {\fft J} \right\} \!\cdot \!{\bf
k}}/2\big).
\nonumber \\
\label{EQ:dens_def}
\end{eqnarray}
The rationale for this generalization goes as follows. We know that in
the undeformed system the probability cloud is asymptotically
isotropic. For an infinitesimal deformation, one might expect the
localization region to be slightly distorted. To lowest order in the
deformation, the matrix characterizing the deformation is ${\fft J}$. 
The other ingredient that
can influence the shape of the localization region is the disorder: 
thus we include a random factor $\eta_{j}(s)$ that weights the departure
of the localization region from spherical symmetry.  
For example, if $\eta_{j}(s)=2$, Eq.~(\ref{EQ:dens_def}) would reduce to
Eq.~(\ref{EQ:affine_den_k_linear}), meaning that the probability cloud
is affinely distorted.  By contrast, if $\eta_{j}(s)=0$
Eq.~(\ref{EQ:dens_def}) would reduce to Eq.~(\ref{EQ:spherical_den_k}),
\ie, the probability cloud would remain spherically symmetric, as it 
is in the undeformed system.  In the same spirit as in the undeformed case,
we assume that the parameters $\eta$ and $\xi$ describing the extent
(and shape) of the fluctuation region are uncorrelated with the
original mean position ${\bf b}$.

By considering $g$ real copies of the system, and adding the
contributions of all monomers, we can explicitly construct the
order parameter of Eq.~(\ref{EQ:opDefinition}):

\eqbreak
\begin{eqnarray}
\Omega_{{\bf k}^{1},\cdots,{\bf k}^{g}} 
& = & (1-q) \,\, \delta_{{\bf k}^{1},{\bf 0}} \times \cdots \times 
\delta_{{\bf k}^{g},{\bf 0}}
\nonumber 
\nopagebreak
\\
& & 
+ q \int \! \frac{d{\bf b}}{V} \,
{\rm e}^{ i{ ({\bf k}^{1} + \cdots + {\bf k}^{g}) \cdot {{\fft S} \cdot 
{\bf b}} } } 
\int \! d{\bf t} 
\int_{0}^{\infty}\!\!\!d\tau 
\int_{-\infty}^{\infty}\!\!\!d\eta\,
\psi({\bf t}, \tau, \eta)\,
{\rm e}^{ i{ ({\bf k}^{1} + \cdots + {\bf k}^{g}) \cdot {\bf t}} }
{\rm e}^{-( 
{{\bf k}^{1}\!\cdot \! \left\{ {\fft I}
		+\eta \, {\fft J} \right\} \!\cdot \!{\bf k}^{1}}
+ \cdots + 
{{\bf k}^{g}\!\cdot \! \left\{ {\fft I}
		+\eta \, {\fft J} \right\} \!\cdot \!{\bf k}^{g}}
)/{2 \tau}}\,.
\label{EQ:op_from_dist_def}
\end{eqnarray}

Here, $\psi({\bf t}, \tau, \eta)$ is the joint statistical 
distribution for the parameters ${\bf t}$, $\tau$ and $\eta$, i.e., 
\beq
\frac{1}{V} \psi({\bf t}, \tau, \eta) \equiv 
\left[ \frac{1}{N}
\sum_{j=1}^{N} \int_{0}^{1} ds\, 
\delta\big({\bf b} - {\bf b}_{j}(s)\big)\, 
\delta\big({\bf t} - {\bf t}_{j}(s)\big)\,
\delta\big(\tau - (\xi_{j}(s))^{-2}\big)\, 
\delta\big(\eta - \eta_{j}(s)\big) 
\right]. 
\label{EQ:psi_def}
\eeq
In order for Eq.~(\ref{EQ:op_from_dist_def}) to reproduce the order 
parameter of Eq.~(\ref{EQ:op_from_dist}) in the limit of zero strain, 
the following
condition on $\psi({\bf t}, \tau, \eta)$ has to be satisfied:
\begin{equation}
\lim_{{\fft S} \to {\fft I}} \psi({\bf t}, \tau, \eta) = 
\delta({\bf t})\,
p(\tau)\, 
\delta(\eta).
\label{EQ:psi_undef}
\end{equation}
The integral over ${\bf b}$ in Eq.~(\ref{EQ:op_from_dist_def})
factorizes for the same reason as in the undeformed system, namely
because ${\bf b}$ is uncorrelated with all the other parameters.

In order to solve the stationary point equations, we need an expression
for $\Omega_{\hat{k}}$, where $\hat{k}$ is a generic replicated wave
vector in $R^{\rm s}$. Obtaining this expression is less
straightforward than in the undeformed case: we have to take into
account the fact that replica $\alpha=0$ is different from all the others
because it is not affected by the deformation. This suggests that for
localized monomers we parametrize the Fourier-transformed individual
particle density by using Eq.~(\ref{EQ:individual_density_gaussian})
for $\alpha=0$ and Eq.~(\ref{EQ:dens_def}) for $\alpha=1,\ldots,n$,
and thus we obtain the following form for $\Omega_{\hat{k}}$:
\bml
\begin{eqnarray}
\Omega_{\hat{k}}
& = & (1-q) \,\, 
\prod\nolimits_{\alpha=0}^{n} 
\delta_{{\bf k}^{\alpha},{\bf 0}} 
\nonumber 
\\
& & 
+ q \int \! \frac{d{\bf b}}{V} \,
{\rm e}^{ i{( {\bf k}^{0} \cdot {\bf b} 
+ \sum_{\alpha=1}^{n} {\bf k}^{\alpha} \cdot {{\fft S} \cdot {\bf b}} ) } }
\int \! d{\bf t} 
\int_{0}^{\infty}\!\!d\tau 
\int_{-\infty}^{\infty}\!\!d\eta\,
\psi({\bf t}, \tau, \eta)\,
{\rm e}^{ i{\big( \sum_{\alpha=1}^{n} {\bf k}^{\alpha} \big) \cdot {\bf t}} }
{\rm e}^{-\big( ({\bf k}^{0})^2 + 
\sum_{\alpha=1}^{n} 
{{\bf k}^{\alpha}\!\cdot \! \left\{ {\fft I}
		+\eta \, {\fft J} \right\} \!\cdot \!{\bf k}^{\alpha}}
\big)/{2 \tau} }
\label{EQ:ord_par_hyp_s}
\\
& = & (1-q)\,\delta_{ {\hat{k}},\hat{0}}
+ q\,\delta_{{\bf k}^{0}
	+{\fft S}\cdot \sum_{\alpha=1}^{n}{\bf k}^{\alpha},
	{\bf 0}}\,
\, \cpo^{\rm s}(\hat{k}).
\label{EQ:ord_par_decomp}
\end{eqnarray}
\eml 
\eqresume
\noindent
To arrive at the second line we have observed that the product of 
wave-vector 
Kronecker 
delta functions corresponds to a delta
function for replicated wave vectors, we have identified the integral
over ${\bf b}$ as a representation of a Kronecker delta function in
wave-vector space, and we have denoted the integral over ${\bf t}$,
$\tau$ and $\eta$ as $\cpo^{\rm s}(\hat{k})$, \ie, the continuous part of
the order parameter in the strained system.

Although it is not trivial to propose a general form for the
probability distribution $\psi({\bf t}, \tau, \eta)$, 
under fairly mild conditions 
it is possible to expand its Fourier transform with
respect to the random displacement ${\bf t}$ to first order in the
strain and to lowest nontrivial order in wave vectors:
\begin{equation}
\int \!\! d{\bf t} \, {\rm e}^{ i{ {\bf p} \cdot {\bf t}} } \, \psi({\bf t},
\tau, \eta) = p(\tau) \, \delta(\eta) + m(\tau,\eta) \, {{\bf p} \!
\cdot \! {\fft J} \! \cdot \! {\bf p}} + {\cal O}({\fft J}^{2}), 
\label{EQ:psi_p_expand}
\end{equation}
with $m(\tau,\eta)$ an unknown function.  The correctness of this
expansion can be justified as follows. The value of the right hand
side of Eq.~(\ref{EQ:psi_p_expand}) in the limit of zero strain is
dictated by
Eq.~(\ref{EQ:psi_undef}). The first order correction in the strain
is determined by assuming that it is invariant under a rotation
of the coordinate system (which is equivalent to a simultaneous
rotation of ${\bf p}$ and ${\fft J}$).  This condition only allows for
the following terms: (i)~a linear function of ${\bf p} \cdot {\fft J}
\cdot {\bf p}$ times any function of ${\bf p}^{2}$ and (ii)~a product
of an invariant linear function of ${\fft J}$ times any function of
${\bf p}^{2}$.  The only quantity linear in ${\fft J}$ and invariant
under rotations is ${\rm tr}\,{\fft J}$, which is zero for
infinitesimal shear strains, as mentioned above.  Thus we only have
term (i), which, to lowest nontrivial order in wave vectors, reduces
to the contribution appearing in Eq.~(\ref{EQ:psi_p_expand}).

The integral over ${\bf t}$ in
Eq.~(\ref{EQ:ord_par_hyp_s}) is the same as that in
Eq.~(\ref{EQ:psi_p_expand}), but with ${\bf p}$ replaced by
\begin{eqnarray}
\sum_{\alpha=1}^{n}{\bf k}^{\alpha} 
& = & - {\fft S}^{-1} \cdot {\bf k}^{0} \nonumber \\
& \approx & - {\bf k}^{0}  
\label{EQ:p_replace}
\end{eqnarray}
The approximation in the second line is consistent with our keeping
only terms linear in the deformation in Eq.~(\ref{EQ:psi_p_expand}).

We are now in the position of being able to simplify the form of
Eq.~(\ref{EQ:ord_par_hyp_s}) substantially, by taking the following
steps: (i) we use Eqs.~(\ref{EQ:psi_p_expand}) and
(\ref{EQ:p_replace}) to perform the integration over the random
displacement ${\bf t}$; (ii)~we expand all terms consistently to
linear order in ${\fft J}$; and (iii)~we define scaling variables in a
way analogous to that shown in Eq.~(\ref{EQ:p_scale}).  
\eqbreak

As a result of these manipulations, we arrive at the following
hypothesis for the continuous part of the order parameter:
\begin{equation}
\cpo^{\rm s}(\hat{k})\!=\!q\!\int_{0}^{\infty}\!\!\!d\theta\,
{\rm e}^{-\hat{k}^2/{\epsilon \theta}}
\Big(\pi(\theta) 
- {\zeta(\theta)\over{\epsilon}}
{\bf k}^{0}\!\cdot \!{\fft J}\!\cdot \!{\bf k}^{0}
- {\wimpf(\theta)\over{\epsilon}}
\sum_{\alpha=1}^{n}
{\bf k}^{\alpha}\!\cdot \!{\fft J}\!\cdot \!{\bf k}^{\alpha}\Big).
\label{EQ:def_hyp}
\end{equation}
Here, $\zeta(\theta)$ and $\wimpf(\theta)$ are new scaling functions,
which describe the change in the continuous part of the order
parameter due to the deformation.  They are unknown at this point, but
they will be determined later by demanding that the
hypothesis (\ref{EQ:def_hyp}) satisfy the stationary point
equations for the deformed system, Eq.~(\ref{EQ:LG_saddle}). 
Note that in Ref.~\cite{REF:Elas_short} the term~(ii) was omitted from 
Eq.~(9); the results, however, are not altered by this omission.

\eqresume

The motivation for the hypothesis, Eq.~(\ref{EQ:def_hyp}), can be
rephrased in a more compact way as follows.  Let us assume that for
small strains $\cpo^{\rm s}(\hat{k})$ is unchanged by a rotation of
the coordinate system (or, equivalently, by simultaneous rotations of
${\fft S}$ and $\hat{k}$). This is evidently true for $\cpo^{\rm
u}(\hat{k})$ (which is a function of $\hat{k}^{2}$). Therefore the
difference between the two quantities $\cpo^{\rm s}(\hat{k})$ and
$\cpo^{\rm u}(\hat{k})$ has the same property.
If we further assume permutation symmetry among replicas
$\alpha=1,\ldots,n$, this difference can only contain, up to lowest
nontrivial order in the deformation and in the wave vectors, the
following terms: (i)~the product of an invariant linear function of
${\fft J}$ and a linear combination of a constant, $({\bf
k}^{0})^{2}$, and $\sum_{\alpha=1}^{n} ({\bf k}^{\alpha})^{2}$; (ii)~a
linear function of ${\bf k}^{0}\cdot {\fft J}\cdot {\bf k}^{0}$; (iii)~a
linear function of $\sum_{\alpha=1}^{n}{\bf k}^{\alpha}\cdot {\fft
J}\cdot {\bf k}^{\alpha}$; and (iv)~a linear function of
$(\sum_{\alpha=1}^{n}{\bf k}^{\alpha} )\cdot {\fft
J}\cdot (\sum_{\beta=1}^{n}{\bf k}^{\beta})$.  The only quantity linear
in ${\fft J}$ and invariant under rotations is ${\rm tr}\,{\fft J}$,
which is zero for infinitesimal shear strains, as mentioned
above. In addition, by using Eq.~(\ref{EQ:p_replace}), any term of type (iv)
is reduced to a term of type (ii). Thus only terms of type (ii) and (iii)
are left, and we recover Eq.~(\ref{EQ:def_hyp}).

\subsection{Solving the stationary-point equations}
We now show that the hypothesis just proposed does indeed satisfy the
stationary point equations in the deformed system, provided that the
gel fraction $q$ and the scaling functions $\pi(\theta)$,
$\zeta(\theta)$, and $\wimpf(\theta)$ satisfy appropriate conditions.

In order to perform the summation over wave vectors in the stationary
point equation, Eq.~(\ref{EQ:LG_saddle}), one has to take into
account the fact that the sum excludes vectors in the one and zero replica
sectors. For any expression $f_{\hat{k}}$ that is zero in the one
replica sector, the following identity is valid in the large volume 
limit: 
\begin{equation}
{\overline{\sum}}_{\hat{k}} f_{\hat{k}} = 
V \int_{\hat{k}} f_{\hat{k}} 
- \lim_{\hat{k} \to \hat{0}} f_{\hat{k}}.
\label{EQ:general_decompose}
\end{equation}
To simplify our notation, we make use of the following shorthand:
\begin{equation}
\int_{\hat{k}} f_{\hat{k}} \equiv V^{n} \int \frac{d\hat{k}}{(2
\pi)^{(1+n)d}} f_{\hat{k}}, 
\label{EQ:notation_k}
\end{equation}
the factor $V^{n}$ in front of the integral will be irrelevant in the
replica limit $n \to 0$, and we will ignore it from now on. Then the
stationary point equation for the deformed system can be rewritten as:
\begin{equation}
0 = 2\Big(3q-\epsilon+\frac{1}{2}|{\hat{k}}|^2\Big) \Omega_{\hat{k}} 
-3 V \int_{\hat{p}} \Omega_{\hat{p}}\, \Omega_{\hat{k}-\hat{p}}.
\label{EQ:saddle_continuous}
\end{equation}
Two observations are in order here. One is technical, namely that the
volume prefactor in the second term, although it might appear
dangerous, is in fact compensated by a factor of $1/V$ coming from the
integrand. The second is more profound, and will be discussed in
detail later: at this point in the argument, the only dependence that
the stationary point equation still has on the deformation, is that
the ``external'' wave vector $\hat{k}$ has to belong to the discrete
set $R^{\rm s}$ rather than $R^{\rm u}$; the other source of
dependence on the deformation, namely the fact that the sum over wave
vectors in the second term was taken for wave vectors restricted to
the discrete set $R^{\rm s}$, has now been eliminated.  \eqbreak

By inserting the hypothesis for the order parameter,
Eqs.~(\ref{EQ:ord_par_decomp}) and~(\ref{EQ:def_hyp}), into the
stationary point condition, Eq.~(\ref{EQ:saddle_continuous}), and
expanding to first order in the strain, we obtain
\begin{eqnarray}
0=\delta_{\tilde{\bf k}^{\rm s},{\bf 0}}\,
&& \left\{ 
2\left(3q^{2}\!-\epsilon q+q
{{\hat{k}}^2}/{2}
\right)
\int_{0}^{\infty}\!\!d\theta\,\pi(\theta)\,
\,{\rm e}^{-\hat{k}^{2}/{\epsilon \theta}}
\right. 
\nonumber \\
&& \qquad \left.
- 3q^{2}\!
\int_{\hat{p}}
\int_{0}^{\infty}\!\!\!d\theta_{1}\,\pi(\theta_{1})
\int_{0}^{\infty}\!\!\!d\theta_{2}\,\pi(\theta_{2})\,
\,{\rm e}^{-\hat{p}^{2}/{\epsilon \theta_{1}}}
\,{\rm e}^{-(\hat{k}-\hat{p})^{2}/{\epsilon \theta_{2}}}
\int \!\! d{\bf m} \, {\rm e}^{i {\bf m}\cdot \tilde{\bf p}^{\rm s} }
\right. 
\nonumber \\
&& \left.
-
2\left(3q^{2}\!-\epsilon q+q
{{\hat{k}}^2}/{2}
\right)
\int_{0}^{\infty}\!\!d\theta\,{\zeta(\theta)\over{\epsilon}}
{\bf k}^{0}\!\cdot \!{\fft J}\!\cdot \!{\bf k}^{0}
\,{\rm e}^{-\hat{k}^{2}/{\epsilon \theta}}
\right. 
\nonumber \\
&& \qquad \left.
+ 6q^{2}\!
\int_{\hat{p}}
\int_{0}^{\infty}\!\!\!d\theta_{1}\, {\zeta(\theta_{1})\over{\epsilon}}
{\bf p}^{0}\!\cdot \!{\fft J}\!\cdot \!{\bf p}^{0}
\int_{0}^{\infty}\!\!\!d\theta_{2}\,\pi(\theta_{2})\,
\,{\rm e}^{-\hat{p}^{2}/{\epsilon \theta_{1}}}
\,{\rm e}^{-(\hat{k}-\hat{p})^{2}/{\epsilon \theta_{2}}}
\int \!\! d{\bf m} \, {\rm e}^{i {\bf m}\cdot \tilde{\bf p}^{\rm s} }
\right. 
\nonumber \\
&& \left.
-
2\left(3q^{2}\!-\epsilon q+q
{{\hat{k}}^2}/{2}
\right)
\int_{0}^{\infty}\!\!d\theta\, {\wimpf(\theta)\over{\epsilon}}
\sum_{\alpha=1}^{n}
{\bf k}^{\alpha}\!\cdot \!{\fft J}\!\cdot \!{\bf k}^{\alpha}
\,{\rm e}^{-\hat{k}^{2}/{\epsilon \theta}}
\right. 
\nonumber \\
&& \qquad \left.
+ 6q^{2}\!
\int_{\hat{p}}
\int_{0}^{\infty}\!\!\!d\theta_{1}\,{\wimpf(\theta_{1})\over{\epsilon}}
\sum_{\alpha=1}^{n}
{\bf p}^{\alpha}\!\cdot \!{\fft J}\!\cdot \!{\bf p}^{\alpha}
\int_{0}^{\infty}\!\!\!d\theta_{2}\,\pi(\theta_{2})\,
\,{\rm e}^{-\hat{p}^{2}/{\epsilon \theta_{1}}}
\,{\rm e}^{-(\hat{k}-\hat{p})^{2}/{\epsilon \theta_{2}}}
\int \!\! d{\bf m} \, {\rm e}^{i {\bf m}\cdot \tilde{\bf p}^{\rm s} }
\right\}. 
\label{EQ:plug_def}
\end{eqnarray}
\noindent
Here we have made use of the notation
\begin{equation}
\tilde{\bf k}^{\rm s} \equiv {\bf k}^{0}
	+{\fft S}\cdot \sum_{\alpha=1}^{n}{\bf k}^{\alpha},
\label{EQ:ktiltil}
\end{equation}
and the integral representation for the Kronecker delta
\begin{equation}
\delta_{{\bf k},{\bf 0}} = \frac{1}{V} \int \!\! d{\bf m} \, {\rm e}^{i
{\bf m}\cdot {\bf k}}.
\label{EQ:Kronecker}
\end{equation}
\noindent
After performing the integrations, first over $\hat{p}$, then over ${\bf
m}$, Eq.~(\ref{EQ:plug_def}) reduces to:
\begin{eqnarray}
0 & = &
\bigg[
2\left(3q^{2}\!-\epsilon q+q
{{\hat{k}}^2}/{2}
\right)
\int_{0}^{\infty}\!\!d\theta\,\pi(\theta)\,
\,{\rm e}^{-\hat{k}^{2}/{\epsilon\theta}}
- 3q^{2}\!
\int_{0}^{\infty}\!\!\!d\theta_{1}\!
\int_{0}^{\infty}\!\!\!d\theta_{2}\,
\pi(\theta_{1}) \pi(\theta_{2})
\,{\rm e}^{-\hat{k}^{2}/{\epsilon(\theta_{1}+\theta_{2})}}
\bigg]
\nonumber \\
& - & \frac{1}{\epsilon} 
\left\{ {\bf k}^{0}\!\cdot \!{\fft J}\!\cdot \!{\bf k}^{0} \right\} 
\bigg[ 
2\left(3q^{2}\!-\epsilon q+q
{{\hat{k}}^2}/{2}
\right)
\int_{0}^{\infty}\!\!d\theta\,\zeta(\theta)\,
\,{\rm e}^{-\hat{k}^{2}/{\epsilon\theta}}
-6q^{2}\!
\int_{0}^{\infty}\!\!\!d\theta_{1}\!
\int_{0}^{\infty}\!\!\!d\theta_{2}\, 
\left(\frac{\theta_{1}}{\theta_{1}+\theta_{2}}\right)^{2} 
\zeta(\theta_{1}) \pi(\theta_{2})
\,{\rm e}^{-\hat{k}^{2}/{\epsilon(\theta_{1}+\theta_{2})}}\bigg]
\nonumber \\
& - &
\frac{1}{\epsilon} \left\{ \sum_{\alpha=1}^{n}
 {\bf k}^{\alpha}\!\cdot \!{\fft J}\!\cdot \!{\bf k}^{\alpha} \right\}
\bigg[ 
2\left(3q^{2}\!-\epsilon q+q
{{\hat{k}}^2}/{2}
\right)
\int_{0}^{\infty}\!\!d\theta\,\wimpf(\theta)\,
\,{\rm e}^{-\hat{k}^{2}/{\epsilon\theta}}
\bigg. 
\nonumber \\
&& \qquad \qquad \qquad \qquad \qquad \bigg.
-6q^{2}\!
\int_{0}^{\infty}\!\!\!d\theta_{1}\!
\int_{0}^{\infty}\!\!\!d\theta_{2}\,
\left(\frac{\theta_{1}}{\theta_{1}+\theta_{2}}\right)^{2} 
\wimpf({\theta}_{1}) \pi(\theta_{2})
\,{\rm e}^{-\hat{k}^{2}/{\epsilon(\theta_{1}+\theta_{2})}}
\bigg]. 
\label{EQ:plug_analog}
\end{eqnarray}
\eqresume

Now, what conditions are forced on the unknown quantities $q$,
$\pi(\theta)$, $\zeta(\theta)$ and $\wimpf(\theta)$ by this stationary
point equation? First, by taking the limit ${\hat{k}}^{2} \to 0$ we
recover the condition for the gel fraction
\begin{equation}
0=-2q\epsilon+3q^2, 
\label{EQ:q_eqn_def}
\end{equation}
which implies that $q = 2\epsilon/3$ for $\epsilon > 0$.  It is not
surprising that we obtain the same gel fraction for the amorphous
solid state as in the unstrained system, as in our motivation for
the order parameter hypothesis we assumed that the monomers that were
localized in the strained system would also be those that were localized
in the unstrained system.

Next, we observe that demanding that Eq.~(\ref{EQ:plug_analog}) be
valid for all $\hat{k} \in R^{\rm s}$ is equivalent to the above
equation for the gel fraction, together with the following
integro-differential equations for the scaling functions
$\pi(\theta)$, $\zeta(\theta)$, and $\wimpf(\theta)$:
\begin{mathletters}
\begin{eqnarray}
\frac{\theta^{2}}{2} \frac{d\pi}{d\theta}
& = & 
(1-\theta)\,\pi(\theta)
-
\phantom{ \frac{2}{\theta^{2}} }
\int_{0}^{\theta} \!\!\!d\theta^{\prime}
\pi(\theta^{\prime})\pi(\theta-\theta^{\prime}), 
\nonumber \\ &&
\label{EQ:old_scpieq}
\\
\frac{\theta^{2}}{2} \frac{d\zeta}{d\theta}
& = &
(1-\theta)\,\zeta(\theta)
-
\frac{2}{\theta^{2}} 
\int_{0}^{\theta} \!\!\!d\theta^{\prime}\,
{\theta'}^{2}\,\zeta(\theta^{\prime})\,\pi(\theta-\theta^{\prime}),
\nonumber \\ &&
\label{EQ:sczetaeq}
\\
\frac{\theta^{2}}{2} \frac{d\wimpf}{d\theta}
\!& = &\!
(1-\theta)\,\wimpf(\theta)
\!-\!
\frac{2}{\theta^{2}} 
\int_{0}^{\theta} \!\!\!d\theta^{\prime}\,
{\theta'}^{2} \wimpf(\theta^{\prime})\,\pi(\theta-\theta^{\prime}).
\nonumber \\ &&
\label{EQ:scphieq}
\end{eqnarray}
\end{mathletters}

As for the boundary conditions satisfied by the scaling functions
$\pi(\theta)$, $\zeta(\theta)$, and $\wimpf(\theta)$, they are
obtained by studying the values of the order parameter in different
regions of $\hat{k}$ space.
First, as noticed in Sec.~\ref{SEC:pi_theta} for the case of
$p(\tau)$, the fact that the order parameter is unity at the origin
(\ie\ $\hat{k} = \hat{0}$) determines the following normalization condition
for $\pi(\theta)$:
\begin{equation}
\int_{0}^{\infty} \!\!d\theta \, \pi(\theta) = 1.
\label{EQ:pinorm}
\end{equation}
Next, to derive boundary conditions for $\zeta(\theta)$ and
$\wimpf(\theta)$ we observe that, from Eq.~(\ref{EQ:opDefinition}),
\begin{equation}
\lim_{ ({\bf k}^{1})^{2}+\cdots+({\bf k}^{g})^{2} \to \infty }
\Omega_{{\bf k}^{1},\ldots,{\bf k}^{g}} = 0, 
\label{EQ:omega_k_infty_phys}
\end{equation}
and consequently that 
\begin{equation}
\lim_{ {\hat{k}}^{2} \to \infty }
\Omega_{\hat{k}} = 0.
\label{EQ:omega_k_infty_rep}
\end{equation}
In order to benefit from this information, we perform the change of
variables 
\begin{equation}
\theta \rightarrow y \equiv \hat{k}^2/{\epsilon \theta}
\label{EQ:y_ket}
\end{equation}
in Eq.~(\ref{EQ:def_hyp}), thus obtaining
\begin{eqnarray}
\lefteqn{ 
\cpo^{\rm s}(\hat{k}) = q\!\int_{0}^{\infty}\!\!\!dy
\,{\rm e}^{-y}
\Big\{\frac{\epsilon}{\hat{k}^2} \Big. \, 
\tilde{\pi}\Big(\frac{\hat{k}^2}{\epsilon y}\Big) 
- \tilde{\zeta}\Big(\frac{\hat{k}^2}{\epsilon y}\Big) \, 
{\bf x}^{0}\!\cdot \!{\fft J}\!\cdot \!{\bf x}^{0} 
} 
\nonumber \\
&& \qquad \qquad \qquad \Big. 
- 
\tilde{\wimpf}\Big(\frac{\hat{k}^2}{\epsilon y}\Big) 
\, \Big. \sum_{\alpha=1}^{n}
{\bf x}^{\alpha}\!\cdot \!{\fft J}\!\cdot \!{\bf x}^{\alpha} \Big\}.
\label{EQ:hyp_y}
\end{eqnarray}
Here, we have defined the functions $\tilde{\pi}(\theta)$,
$\tilde{\zeta}(\theta)$, and $\tilde{\wimpf}(\theta)$ by 
\begin{equation}
\tilde{\pi}(\theta) \equiv \theta^{2} \pi(\theta), \qquad 
\tilde{\zeta}(\theta) \equiv \theta^{2} \zeta(\theta), \qquad 
\tilde{\wimpf}(\theta) \equiv \theta^{2} \wimpf(\theta), 
\label{EQ:tilde_functions}
\end{equation}
and the unit vector $\hat{x} = \{ {\bf x}^{0},\ldots,{\bf x}^{n} \}$ by
\begin{equation}
{\bf x}^{\alpha} \equiv \frac{ {\bf k}^{\alpha} }{ \sqrt{\hat{k}^2} },
 \qquad \qquad (\alpha= 0,\ldots,n).
\label{EQ:x_unit}
\end{equation}
{}From the expression~(\ref{EQ:hyp_y}) for the order parameter
hypothesis, and the exponential decay of $\pi(\theta)$ for $\theta \to
\infty$, it follows that
\begin{equation}
\lim_{ {\hat{k}}^{2} \to \infty }
\Omega_{\hat{k}} 
= - \lim_{\theta \to \infty} 
\Big(\tilde{\zeta}(\theta) \,{\bf x}^{0}\!\cdot \!{\fft J}\!\cdot \!{\bf x}^{0}
+ \tilde{\wimpf}(\theta)  \! \sum_{\alpha=1}^{n}
{\bf x}^{\alpha}\!\cdot \!{\fft J}\!\cdot \!{\bf x}^{\alpha} \Big).
\label{EQ:lim_tilde}
\end{equation}
However, this limit must be zero regardless of the direction of
$\hat{x}$, and consequently we obtain the following boundary
conditions for $\zeta(\theta)$ and $\wimpf(\theta)$:
\bml
\begin{eqnarray}
\lim_{\theta \to \infty} \theta^{2} \zeta(\theta) & = &  0,
\label{EQ:zeta_bound} \\ 
\lim_{\theta \to \infty} \theta^{2} \wimpf(\theta) & = &  0.  
\label{EQ:phi_bound}
\end{eqnarray}
\eml
To obtain boundary conditions at $\theta = 0$, one only needs to examine
the integro-differential equations (\ref{EQ:old_scpieq}),
(\ref{EQ:sczetaeq}), and (\ref{EQ:scphieq}) themselves. Near the origin, the
integral terms can be neglected, and all three equations reduce to the
form:
\begin{equation}
\frac{\theta^{2}}{2} \frac{df}{d\theta}
= (1-\theta)\,f(\theta),
\label{EQ:eq_pi_linear}
\end{equation}
where $f$ stands for $\pi$, $\zeta$, or $\wimpf$. This is a 
first order linear differential equation having the solution 
\begin{equation}
f(\theta) = A \frac{{\rm e}^{-2/\theta}}{{\theta}^{2}},
\label{EQ:pi_linear}
\end{equation}
with $A$ an arbitrary constant. Consequently, all three scaling
functions vanish rapidly at the origin. 

As the reader has probably already noticed, the integro-differential
equations and the boundary conditions that apply to both
$\zeta(\theta)$ and $\wimpf(\theta)$ are linear and homogeneous. 
This implies that one of two possibilities must hold for each one of
these functions: either it is identically zero, or it is only
determined up to an arbitrary multiplicative constant. [By contrast,
in the case of $\pi(\theta)$, the integro-differential
equation~(\ref{EQ:old_scpieq}) is nonlinear, and the condition of
Eq.~(\ref{EQ:pinorm}) is linear but inhomogeneous, and the scale of
the solution is well determined.]\thinspace\  The latter possibility
does not seem to be easy to justify on physical grounds, as it would
imply that the stationary-point equations leave the order parameter
undetermined. In fact, if this were the case, there would be a
continuous family of order parameters such that the continuous parts
$\cpo^{\rm s}(\hat{k})$ for members of the family differ to varying
degrees from the continuous part of the order parameter corresponding
to the amorphous solid state of the unstrained system. One could,
however, imagine that we are missing 
some additional physical constraint that
fixes the scale of these two functions, and therefore the above
argument is suggestive but not conclusive. To settle the issue of
which of the two possibilities holds for $\zeta(\theta)$ and
$\wimpf(\theta)$, we show, in Appendix~\ref{APP:z_0}, by analytic
manipulation of the integro-differential equations and boundary
conditions, that both $\zeta(\theta)$ and $\wimpf(\theta)$ are
identically zero.

The fact that both $\zeta(\theta)$ and $\wimpf(\theta)$ are
identically null implies the {\it a priori\/} most 
surprising result of this paper: the continuous part of the order 
parameter {\em does not change\/} to first order in the strain, 
\ie, $\cpo^{\rm s}(\hat{k})=\cpo^{\rm u}(\hat{k})$. 
This conclusion is consistent with the phantom network 
picture~\cite{REF:phantom,REF:MarkErman}.
It also suggests that $\cpo^{\rm s}(\hat{k})=\cpo^{\rm u}(\hat{k})$
for finite (and not merely infinitesimal) deformations.  Indeed, our
order-parameter hypothesis turns out to satisfy the stationary-point
equation for arbitrarily strained systems. 

To see this, let us return to the stationary point equation
(\ref{EQ:saddle_continuous}). As was mentioned earlier,
Eq.~(\ref{EQ:saddle_continuous}) applies both for the unstrained and
for the strained systems, the only difference between the two cases
being that in the unstrained case the ``external'' replicated wave
vector $\hat{k}$ belongs to the discrete set $R^{\rm u}$, whereas in
the strained case $\hat{k}$ belongs to the set $R^{\rm s}$. By
inserting the form for the order parameter given by
Eq.~(\ref{EQ:ord_par_decomp}), but now with $\cpo^{\rm
s}(\hat{k})=\cpo^{\rm u}(\hat{k})$ [\ie, given by
Eq.~(\ref{EQ:def_hyp}) with $\zeta(\theta) = \wimpf(\theta) = 0$] we
find that the stationary point equation is satisfied provided
$\pi(\theta)$ satisfies Eq.~(\ref{EQ:scpieq}).

One way of understanding this result is to consider that in order for
the shape of the fluctuation region to be affected by the externally
imposed strain, this strain has to be somehow communicated to the
individual monomers. This is most likely the effect of the deformation
of the \lq\lq cage \rq\rq of surrounding polymers that form the local
environment 
at each point. However, when the interlocking of loops is neglected,
as in the present calculation, this \lq\lq cage \rq\rq exerts no
effect. Therefore, this result should be taken with caution, as its
validity might not extend beyond the region near the transition, in
which the approximation of neglecting the interlocking of loops is
fully justified.

One might be tempted at this point to assume that the order parameter
is completely unchanged when the system is deformed. However, this is
not quite correct.  In addition to the stationary point equation, the order
parameter has to satisfy the boundary conditions in real space for the
deformed system. This means that the hypothesis of
Eq.~(\ref{EQ:ord_par_decomp}) for $\Omega_{\hat{k}}$ is physically
meaningful only for $\hat{k}$ belonging to the set of allowed
replicated wave vectors $R^{\rm s}$. If the order parameter
corresponding to the unstrained system were retained, there would be a
factor $\delta_{\tilde{\bf k}, {\bf 0}}$ in the term corresponding to
the localized monomers that would be zero for generic values of the
deformation matrix $\fft S$ unless {\em both\/} ${\bf k}^{0} = {\bf 0} $
and $\sum_{\alpha=1}^{n} {\bf k}^{\alpha} = {\bf 0} $. As in the
undeformed system this same factor is nonzero for $\sum_{\alpha=1}^{n}
{\bf k}^{\alpha} = - {\bf k}^{0} \neq {\bf 0} $, this would give rise
to an unphysical discontinuity in the order parameter as a function of
the deformation. On the other hand, the modified delta factor
$\delta_{\tilde{\bf k}^{\rm s}, {\bf 0}}$ that appears in
Eq.~(\ref{EQ:ord_par_decomp}) takes into account the shift in the
reciprocal lattice due to the deformation, and displays no such
discontinuity.

\subsection{Change in free energy with deformation; shear modulus}
\label{SEC:free_en_chan}
We now have all the ingredients necessary to calculate the change in 
the free energy $\Delta f$, to leading order in $\epsilon$, due to 
the deformation of the system:
\begin{equation}
\Delta f = 
d\lim_{n\rightarrow 0} 
\left\{
{\cal F}^{\rm s}_{n}\big(\{\Omega^{\rm s}_{\hat{k}}\}\big)
\!-\!
{\cal F}_{n} \big(\{\Omega^{\rm u}_{\hat{k}}\}\big)\right\}.
\label{EQ:change_def}
\end{equation}
Here $\Omega^{\rm s}_{\hat{k}}$ and $\Omega^{\rm u}_{\hat{k}}$ are,
respectively, the stationary-point values of the order parameter for the
strained and unstrained systems. Similarly, ${\cal F}^{\rm s}_{n}$ and
${\cal F}_{n}$ respectively denote the free energy functionals for the
strained and unstrained systems.  
As we show in App.~\ref{APP:FE_change}, the
free-energy change due to the deformation is 
\begin{equation}
\Delta f  = 
\frac{2}{27} \epsilon^{3}\,\,{\rm tr}\,\,
({\fft S}^{2}- {\fft I}).
\label{EQ:total_change}
\end{equation}
Thus we can extract the value of the
static shear modulus $\shmd$ of the amorphous solid state near the
solidification transition (with physical units restored):
\begin{equation}
\shmd=k_{\rm B}T\,N\,C\,\epsilon^{3},
\label{EQ:shmd_result}
\end{equation}
where $k_{\rm B}$ is Boltzmann's
constant, $T$ is the temperature, and $C$ is a model-dependent
positive constant.  Hence, we see that the static shear modulus near
the vulcanization transition is characterized by the exponent
$\shex=3$, in agreement with the classical
result~\cite{REF:DobGor,REF:PGDGbook}.  A simple scaling argument,
viz., that the modulus should scale as two powers of the order
parameter ($q^2$) and two powers of the gradient ($\xi^{-2}_{\rm
typ}$), leads to the same value for the exponent $\shex$.

\section{Concluding remarks}
\label{SEC:conclusions}
In this Paper we have presented a microscopic derivation of the
static elastic response of a system of randomly crosslinked
macromolecules near the amorphous solidification transition. 

{}From the technical point of view, we have modeled the deformation of
the system by changing the boundary conditions in real space.
A point that
required special care was how to include in our formulation the
physical information that the system had been crosslinked {\em before\/}
it was deformed. This results in an asymmetry in the replica
formulation of the problem: in the case we are studying, replica
$\alpha = 0$ describes the system {\em before\/} the deformation is
applied, and replicas $\alpha=1,\ldots,n$ describe the system in its
actual state of deformation. 

The physical picture that emerges from the results of this Paper has
the following features: (i)~the amorphous solid state, which had been 
previously shown
to be characterized structurally by the
localization of a nonzero fraction of particles, is also characterized
by having a nonzero static shear modulus; (ii)~the static shear
modulus scales as the third power of the excess crosslink density
(beyond its value at the transition)
\cite{FNOTE:previous}; and (iii)~the form of localization
exhibited by the particles is left unchanged by the strain.  

A possible explanation for the spherical localization regions that the
particles exhibit even under externally applied stress  
might be that in the regime near the transition most monomers in the
infinite cluster are very loosely connected, and thus their behavior
is dominated by the maximization of entropy, which is obtained by
allowing them to fluctuate in all directions.  It is not implausible
that strain-induced changes in the pattern of localization would
emerge from a more detailed analysis of the effects of the
excluded-volume interaction, at least at higher crosslink
densities. This is because at higher crosslinks densities, the
macromolecular network is more tightly bound, and the topological
barriers generated by interlocking of macromolecular loops are more
significant.

Finally, let us point out that since the treatment presented here only
depends on the form of the free-energy
functional~\cite{REF:landau} near the transition, and not any specific
semi-microscopic model, the approach to elasticity described here
should be generally applicable not only to systems of randomly
crosslinked flexible macromolecules, but also to other equilibrium
amorphous solid forming systems.

\section{Acknowledgements} 
\label{SEC:acknowledgments}
We thank 
S.~Barsky, B.~Jo{\'o}s, M.~Plischke, W.~Peng and A.~Zippelius
for stimulating discussions.  We gratefully acknowledge support
from National Science Foundation grants 
DMR94-24511 (HEC, PMG) and 
DMR99-75187 (PMG).

\appendix

\section{Correction to the order parameter under strain} 
\label{APP:z_0}
In this Appendix we show that the only solution to
Eq.~(\ref{EQ:sczetaeq}) that satisfies the boundary condition
Eq.~(\ref{EQ:zeta_bound}) is the null function 
$\zeta(\theta) \equiv 0$ for all $\theta$. 
Our approach is to assume that a nonzero solution exists, and then to 
arrive at a contradiction.  As the equations and boundary conditions 
are identical for $\zeta(\theta)$ and $\wimpf(\theta)$, showing that 
$\zeta(\theta)$ is identically null would imply that the same holds 
for $\wimpf(\theta)$.

It is convenient to work with $\tilde{\zeta}(\theta)$ instead of
$\zeta(\theta)$. In terms of $\tilde{\zeta}(\theta)$, the
integro-differential equation reads:
\beq
\frac{\theta^{2}}{2} \frac{d\tilde{\zeta}}{d\theta}
= 
\tilde{\zeta}(\theta)
-
2 \int_{0}^{\theta} \!\!\!d\theta^{\prime}\,
\tilde{\zeta}(\theta^{\prime})\,\pi(\theta-\theta^{\prime}).
\label{EQ:sctzeq}
\eeq
The boundary condition is simply 
\beq
\lim_{\theta \to \infty} \tilde{\zeta}(\theta) = 0.
\label{EQ:bound_tz}
\eeq
It turns out that it is possible to derive a simple differential
equation for the Laplace transform $\hat{\varrho}(s)$ of the function 
\beq
\varrho(\theta) \equiv \frac{d\tilde{\zeta}}{d\theta}.
\label{EQ:def_der}
\eeq 
By starting with Eq.~(\ref{EQ:sctzeq}), and using properties of 
the Laplace transform, one obtains (after some algebra) the equation
\beq
\frac{d^{2}\hat{\varrho}(s)}{ds^{2}} = \frac{2}{s} \hat{\varrho}(s) \big(1
- \hat{\pi}(s) \big), 
\label{EQ:lap_eq}
\eeq
and the boundary condition
\beq
\hat{\varrho}(0) = \int_{0}^{\infty} \!\!\!d\theta\,
\frac{d\tilde{\zeta}}{d\theta} = \lim_{\theta \to \infty}
\tilde{\zeta}(\theta)  - \tilde{\zeta}(0) = 0.
\label{EQ:lap_bou}
\eeq
The function $\hat{\pi}(s)$ appearing in Eq.~(\ref{EQ:lap_eq}) is the
Laplace transform 
of the scaled
probability distribution $\pi(\theta)$ for the unstrained system. By
using its expansion for small $s$, namely
\beq
\hat{\pi}(s) = 1 - s \langle \theta \rangle_{\pi} + {\cal O}(s^{2}),
\label{EQ:hatpi_small}
\eeq
one can immediately show that Eq.~(\ref{EQ:lap_eq}) has a regular singular
point at the origin, and thence use the Frobenius method~\cite{REF:Bender}
to obtain the asymptotic forms near the origin of two linearly
independent solutions:
\bml
\bea
\hat{\varrho}_{1}(s) & = & s - s^{2} + {\cal O}(s^{3}), 
\label{EQ:sol1} \\
\hat{\varrho}_{2}(s) & = & \frac{1}{2} - s \ln{s} + {\cal O}(s).
\label{EQ:sol2}
\eea%
\eml%
Any solution of Eq.~(\ref{EQ:lap_eq}) can be written as a linear
combination of these two. Due to the boundary
condition~(\ref{EQ:lap_bou}), the coefficient of $\hat{\varrho}_{2}(s)$
must be zero. Therefore $\hat{\varrho}(s)$ is some real multiple of 
$\hat{\varrho}_{1}(s)$.

We have not been able to integrate Eq.~(\ref{EQ:lap_eq}) analytically.
However, it is straightforward to integrate it numerically, using the
behavior given by Eq.~(\ref{EQ:sol1}) as the initial condition. The
numerical solution thus obtained diverges at infinity; but as
$\hat{\varrho}(s)$ is the Laplace transform of a function, 
it goes 
to zero at infinity. Therefore, by assuming that a nonzero solution
can be found satisfying both Eq.~(\ref{EQ:sczetaeq}) and
Eq.~(\ref{EQ:zeta_bound}), we have arrived at a contradiction.

\section{Free energy change under strain} 
\label{APP:FE_change}

We need to compute the difference between the free energy 
of the deformed system, 
${\cal F}^{\rm s}_{n}\big(\{\Omega^{\rm s}_{\hat{k}}\}\big)$, 
and the undeformed system, 
${\cal F}_{n}\big(\{\Omega^{\rm u}_{\hat{k}}\}\big)$, 
as a function of the deformation matrix ${\fft S}$. 
{}From Eq.~(\ref{EQ:free_en_def}) we see that 
${\cal F}^{\rm s}_{n}\big(\{\Omega^{\rm s}_{\hat{k}}\}\big)$ 
contains both a
quadratic and a cubic term in $\Omega^{\rm s}_{\hat{k}}$.
We first study the quadratic term. We make use of
Eq.~(\ref{EQ:general_decompose}) to write, in the large volume limit,
\begin{eqnarray}
&&{\overline{\sum}}_{\hat{k} \in  R^{\rm s} }
\Big(-\epsilon+ 
\frac{|\hat{k}|^2}{2}\Big)
\big\vert\Omega_{\hat{k}}\big\vert^{2} \\
&&= \epsilon \lim_{\hat{k} \to \hat{0}}
\big\vert\Omega_{\hat{k}}\big\vert^{2} + 
V \int_{\hat{k}} \Big(-\epsilon+
\frac{|\hat{k}|^2}{2}\Big)
\big\vert\Omega_{\hat{k}}\big\vert^{2}. 
\label{EQ:quadratic_decompose}
\end{eqnarray}
The term associated with the limit $\hat{k} \to \hat{0}$ has the value
$\epsilon q^2$, independent of ${\fft S}$, and is thus irrelevant for
the present purposes. We concentrate on computing the integral
\begin{equation}
I \equiv V \int_{\hat{k}} \Big(-\epsilon+\frac{|\hat{k}|^2}{2}\Big)
\big\vert\Omega_{\hat{k}}\big\vert^{2}. 
\label{EQ:int_quad}
\end{equation}

To make the analysis more digestible, we define the notations
\begin{equation}
\int_{\theta} {\cdots} \: \equiv \int_{0}^{\infty}\!\!\!d\theta\,
{\cdots}\,\pi(\theta)\,
 \quad \mbox{and} \quad a \equiv
\frac{2}{\epsilon}\Big(\frac{1}{\theta_1}+\frac{1}{\theta_2}\Big). 
\label{EQ:notation_theta}
\end{equation}

The first step is to insert the form of the order parameter for the
solid phase, Eqs.~(\ref{EQ:ord_par_decomp}) and~(\ref{EQ:def_hyp}), and
use the fact that $\zeta(\theta) = \wimpf(\theta)\equiv 0$. 
We then have 
\begin{eqnarray}
\lefteqn{I = 
V \int_{\hat{k}} \Big(-\epsilon+\frac{|\hat{k}|^2}{2}\Big)
\left(q \, \delta_{{\tilde{\bf k}}^{\rm s}, {\bf 0}} \int_{\theta}
{\rm e}^{-\hat{k}^2/{\epsilon \theta}} \right)^{2} }
\\
& = & 
q^{2} \int_{\theta_1} \!\int_{\theta_2} \!\int_{\hat{k}}
\Big(-\epsilon+\frac{|\hat{k}|^2}{2}\Big) {\rm e}^{-a \hat{k}^2/2}
\nonumber \\
& & \qquad \qquad \times \int \!\!d{\bf m}\, {\rm e}^{i{\bf m}\cdot ({\bf
k}^{0} + {\fft S}\cdot \sum_{\alpha=1}^{n}{\bf k}^{\alpha})} 
\nonumber \\ 
& = & 
q^{2} V^{n} \!\int_{\theta_1} \!\int_{\theta_2}
\!\Big(\!-\epsilon-\frac{d}{da}\Big) 
\!\int \!\!d{\bf m}\, \frac{{\rm e}^{-m^2/2a}}{(2 \pi a)^{d/2}}
\Big(\frac{{\rm e}^{-({\fft S}m)^2/2a}}{(2 \pi a)^{d/2}}\Big)^{n}  
\nonumber \\  
& = & 
q^{2} V^{n} \!\int_{\theta_1} \!\int_{\theta_2}
\!\Big(\!-\epsilon-\frac{d}{da}\Big)
\left\{ (2 \pi a)^{nd} [\det({\fft I} + n {\fft S}^{2})]
\right\}^{-1/2}   \nonumber \\
& = & 
q^{2} (1+n\ln{V}) (-\epsilon+{\cal O}(n)) \Big(1 - \frac{n}{2} {\rm tr}\,({\fft
S}^{2})\Big) + {\cal O}(n^{2}), \nonumber \\
&&
\label{EQ:quadratic_result}
\end{eqnarray}
where we have only kept the lowest two powers of the number $n$ of replicas in
the result.
The change in this term due to the deformation is 
\begin{equation}
\Delta I = \frac{n}{2} \epsilon q^{2} {\rm tr}\,({\fft S}^{2} - {\fft
I} ) + {\cal O}(n^{2}). 
\label{EQ:quadratic_change}
\end{equation}

Now, to compute the cubic term, we use Eq.~(\ref{EQ:general_decompose})
repeatedly to obtain 
\begin{eqnarray}
&&
-\,{\overline{\sum}}_{{\hat{k}_1}{\hat{k}_2}{\hat{k}_3} \in  R^{\rm s} }
\Omega_{\hat{k}_1}\,
\Omega_{\hat{k}_2}\,
\Omega_{\hat{k}_3}\,
\delta_{{\hat{k}_1}+{\hat{k}_2}+{\hat{k}_3}, {\hat{0}}}\, \nonumber \\
&& 
=
- V^{2} \!\int_{\hat{k}_1} \!\int_{\hat{k}_2} \Omega_{\hat{k}_1}\,
\Omega_{\hat{k}_2}\, \Omega_{-\hat{k}_1-\hat{k}_2}\, 
+ 3 V q \!\int_{\hat{k}}
\big\vert\Omega_{\hat{k}}\big\vert^{2} 
- 2 q^{3}. \nonumber \\
&&
\label{EQ:cubic_decompose}
\end{eqnarray}
Next, by inserting the form (Eqs.~\ref{EQ:ord_par_decomp}
and~\ref{EQ:def_hyp}) of the order parameter, the first term on the
right hand side yields
\begin{eqnarray}
&& 
J \equiv -q^{3} \!\int_{\theta_1} \!\int_{\theta_2} \!\int_{\theta_3}
\!\int_{\hat{k}_1} \!\int_{\hat{k}_2}  
{\rm e}^{-{\hat{k}_1}^2/{\epsilon \theta_1}-{\hat{k}_2}^2/{\epsilon
\theta_2}-({\hat{k}_1}+{\hat{k}_2})^2/{\epsilon \theta_3} } 
\nonumber \\
&&
\;\times   
\!\int \!\!d{\bf m}_{1}\, {\rm e}^{i{\bf m}_{1}\cdot ({\bf k}_{1}^{0} +
{\fft S}\cdot \sum_{\alpha=1}^{n}{\bf k}_{1}^{\alpha})} 
\!\int \!\!d{\bf m}_{2}\, {\rm e}^{i{\bf m}_{2}\cdot ({\bf k}_{2}^{0} +
{\fft S}\cdot \sum_{\alpha=1}^{n}{\bf k}_{2}^{\alpha})} 
\nonumber \\
&&
=
-q^{3} V^{2n}\!
\int\limits_{\theta_1}\!\!
\int\limits_{\theta_2}\!\!
\int\limits_{\theta_3}\!
\int\!\!\!
d{\bf m}_{1}\,\!\!
\int\!\!
d{\bf m}_{2}\, 
\frac{ \exp\Big[\!-\frac{1}{2} ({\bf m}_{1} {\bf m}_{2}) {\fft A}\!\left({\bf
m}_{1} \atop {\bf m}_{2} \right) \Big] } 
{(4 \pi^{2} \det{\fft A})^{d/2}} \nonumber \\
&&
\quad
\times 
\left(\frac{ \exp\Big[ -\frac{1}{2} ({\fft S}{\bf m}_{1} {\fft S}{\bf
m}_{2}) {\fft A} \left({\fft S}{\bf m}_{1} \atop {\fft S}{\bf
m}_{2} \right) \Big] }  
{(4 \pi^{2} \det{\fft A})^{d/2}} \right)^{n}, 
\label{EQ:cubic_A}
\end{eqnarray}
where the $2 \times 2$ matrix ${\fft A}$ is the inverse of the matrix
\begin{equation}
\frac{2}{\epsilon}
\pmatrix{
\frac{1}{\theta_{1}} + \frac{1}{\theta_{3}} & \frac{1}{\theta_{3}} \cr
\frac{1}{\theta_{2}} & \frac{1}{\theta_{2}} + \frac{1}{\theta_{3}} \cr
	}\label{EQ:matrix_A}
\end{equation}
By performing the Gaussian integration over 
${\bf m}_{1}$ and 
${\bf m}_{2}$,
and expanding in powers of $n$, we obtain 
\begin{eqnarray}
&&
J =
-q^{3} V^{2n} \!\int_{\theta_1} \!\int_{\theta_2} \!\int_{\theta_3} 
(4 \pi^{2} \det{\fft A})^{-nd/2} 
[\det({\fft I} + n {\fft S}^{2})]^{-1} 
\nonumber \\
&&
=
-q^{3} 
(1+2n\ln{V}) (1 - n {\rm tr}\,({\fft S}^{2})) 
(1+{\cal O}(n)) + {\cal O}(n^{2}).
\nonumber \\
&&
\label{EQ:cubic_result}
\end{eqnarray}
For this term, the change due to the deformation is
\begin{equation}
\Delta J = n q^{3} {\rm tr}\,({\fft S}^{2} - {\fft I} ) +
{\cal O}(n^{2}). 
\label{EQ:cubic_change}
\end{equation}
Similarly, the second term on the right hand side of
Eq.~(\ref{EQ:cubic_decompose}) can be evaluated to yield
\begin{eqnarray}
&& K = 
3 V q \int_{\hat{k}}
\big\vert\Omega_{\hat{k}}\big\vert^{2} \nonumber \\
&&
=
3 q^{3} (1+n\ln{V}) (1 - \frac{n}{2} {\rm tr}\,({\fft S}^{2}))  
(1+{\cal O}(n)) + {\cal O}(n^{2}), \nonumber \\
&&
\label{ER:cubic_second_result}
\end{eqnarray}
and its change under deformation is 
\begin{equation}
\Delta K = -\frac{3n}{2} q^{3} {\rm tr}\,({\fft S}^{2} -
{\fft I} ) + {\cal O}(n^{2}). 
\label{EQ:cubic_second_change}
\end{equation}
By combining the contributions given in Eqs.~(\ref{EQ:quadratic_change}),
(\ref{EQ:cubic_change}), and (\ref{EQ:cubic_second_change}), dividing
by the number $n$ of replicas, and taking
into account the fact that $q = 2\epsilon/3$, we obtain the
free-energy change due to the deformation given in 
Eq.~(\ref{EQ:total_change}).


\end{multicols}

\begin{references}
\bibitem{FNOTE:MF_DG} 
In the vulcanization of macromolecular melts, the width of the
critical region, in which mean-field theory fails, is of the order of
$L^{(d-2)/(d-6)}$ (where $L$ is the length of a macromolecule), \ie,
it vanishes in the limit of long macromolecules for dimension $d >
2$~\cite{REF:PGDGbook,REF:Loop_Exp}.
\bibitem{FNOTE:notation}
As is conventional in statistical mechanics, we use the same notation,
$\Omega$, for the equilibrium value of the order parameter
and for the fluctuating variable (of which it is the expectation value
and which features in the Landau free energy). Which meaning is implied
can be readily ascertained from the context.
\bibitem{REF:Erdos}
In the context of graph theory it has been 
shown by Erd{\H o}s and R{\'e}nyi~\cite{REF:ER} that for a random 
graph of $N$ points and $\mu ^{2}N/2$ edges the probability 
for the fraction of points in the largest component to differ 
from the solution $q$ of Eq.~(\ref{EQ:scforQ}) 
vanishes in the $N\rightarrow\infty$ limit.
\bibitem{FNOTE:other_solutions}
One could speculate that the solution to the stationary-point equations is
not unique, and that both a solution dominated by long localization
lengths and a second solution dominated by shorter localization
lengths are present. Even in that case, the physical argument already
given would strongly suggest that the ``long-localization-length''
solution is the physically relevant one.  Besides, to the best of our
knowledge, no second solution has been found, despite attempts by
the authors and by others.
\end{references}
\end{document}